\documentclass[pra,twocolumn,showpacs,superscriptaddress,notitlepage]{revtex4-2}
\usepackage{makeidx}
\usepackage[utf8]{inputenc}
\usepackage{graphicx}
\usepackage{amsmath}

\DeclareMathOperator{\sech}{sech}
\usepackage{mathrsfs}
\usepackage{graphicx}
\usepackage{braket}
\usepackage[subrefformat=parens,labelformat=parens,caption=false]{subfig}
\usepackage{amsfonts}
\usepackage{amssymb}
\usepackage{dcolumn}
\usepackage{bm}
\usepackage{bbm}
\usepackage{color}
\usepackage[dvipsnames]{xcolor}
\usepackage{esint}
\usepackage{lineno}
\usepackage{adjustbox}
\usepackage{verbatim}
\usepackage{cancel}
\usepackage{multirow}

\usepackage[breaklinks,
            colorlinks,
            urlcolor=blue,
            linkcolor=blue,
            anchorcolor=blue,
            citecolor=blue]{hyperref}
           
\usepackage{soul}

\newcommand{\dmtwo}[1]{{\color{black} #1}}

\begin{document}

\title{Stability of macroscopic spin ensembles against inhomogeneous dephasing}

\author{Wai-Keong Mok}
\affiliation{Institute for Quantum Information and Matter, California Institute of Technology, Pasadena, CA 91125, USA}
\affiliation{Centre for Quantum Technologies, National University of Singapore, 3 Science Drive 2, Singapore 117543}

\author{Leong-Chuan Kwek}
\affiliation{Centre for Quantum Technologies, National University of Singapore, 3 Science Drive 2, Singapore 117543}
\affiliation{MajuLab, CNRS-UNS-NUS-NTU International Joint Research Unit, Singapore UMI 3654, Singapore}
\affiliation{National Institute of Education, Nanyang Technological University, Singapore 637616, Singapore}
\affiliation{Quantum Science and Engineering Centre (QSec), Nanyang Technological University, Singapore}

\author{Steven Touzard}
\affiliation{Centre for Quantum Technologies, National University of Singapore, 3 Science Drive 2, Singapore 117543}
\affiliation{MajuLab, CNRS-UNS-NUS-NTU International Joint Research Unit, Singapore UMI 3654, Singapore}
\affiliation{Department of Materials Science and Engineering, National University of Singapore, Singapore}
\affiliation{Department of Physics, National University of Singapore, Singapore}

\begin{abstract}
   Spin ensembles play a pivotal role in various quantum applications such as metrology and simulating many-body physics. Recent research has proposed utilizing spin cat states to encode logical quantum information, with logical lifetimes potentially on the order of seconds, achieved via enhanced collective interactions that scale with system size. We investigate the dynamics of spin cat states under inhomogeneous broadening, revealing a phenomenon termed `parity-sensitive inhomogeneous dephasing': for small amplitudes, odd cat states are significantly more susceptible to inhomogeneous dephasing compared to even cat states due to the difference in parity symmetry. This discrepancy between even and odd cat states vanishes at large amplitudes, and behave similarly to a spin coherent state with the same amplitude. To analyze the stability of the spin coherent state, we perform a mean-field analysis of the driven-dissipative dynamics, from which we identify a synchronization phase transition wherein the ensemble becomes completely dephased beyond a critical inhomogeneous linewidth. The mean-field analysis suggests that the dissipative stabilization can suppress the decoherence effects from inhomogeneous broadening. We argue that the stability of the mean-field model provides a reasonable estimate for that of spin cat states with a large amplitude in the full quantum model. Our findings shed light on the stability of collective spin states, important for advancing quantum technologies.            
\end{abstract}
\maketitle

\section{Introduction}
Coherent manipulation of spin ensembles is crucial for scalable implementations of many quantum technologies, such as quantum metrology~\cite{Pezze2018Quantum,Zhou2020quantum,Arunkumar2023quantum}, computation~\cite{Wesenberg2009quantum,Brion2007quantum,Barrett2010scalable} and simulation~\cite{Georgescu2014quantum,Islam2011onset}. In particular, spin ensembles can be engineered to behave collectively as a single high-dimensional quantum system. This allows for enhanced precision in quantum sensors~\cite{Koppenhofer2022dissipative}, and is also very useful in quantum repeater protocols~\cite{Duan2001long,Sangouard2011quantum}. Collective spin dynamics has also led to the discovery of interesting physical phenomena such as Dicke superradiance~\cite{Dicke1954coherence,Gross1982superradiance}, which remains an active field of study in both theoretical~\cite{Xu2014synchronization,Masson2020many,Masson2022universality,Sierra2021dicke,Robicheaux2021theoretical,Malz2022large,Silvia2023many,Mok2023dicke} and experimental~\cite{Raino2020superradiant,Raino2018superfluorescence,Lei2023many} frontiers to harness it for various physical applications.

Recently, it was proposed that collective spin states are potentially useful for encoding logical quantum information. Motivated by the experimental success of bosonic cat qubits~\cite{Leghtas2015confining,Lescanne2020exponential} for quantum error correction in superconducting circuits, the idea of spin cat qubits is based on encoding the logical qubit in macroscopic superpositions of spin coherent states~\cite{Radcliffe1971some}, i.e., cat states. The spin cat states are dissipatively stabilized by engineering collective two-body losses in the spin ensemble~\cite{qin2021generating}, analogous to the protocol developed for bosonic systems. For realistic experimental parameters, it was estimated in Ref.~\cite{qin2021generating} that the spin cat qubit has a lifetime on the order of seconds, several orders of magnitude larger than the state-of-the-art lifetimes for bosonic cat qubits. This substantial improvement fundamentally stems from the enhanced collective interactions in the spin ensemble which scales as $\sqrt{N}$, where $N$ is the system size. 

In this work, we study the robustness of spin cat states in the presence of inhomogeneous broadening. This can arise for example from Doppler shifts in atomic gas clouds or spatial inhomogeneity in the electric or magnetic fields in solid state systems. Such imperfections break the permutation symmetry of the spin system, which inhibits the collective behavior. We consider the quantum driven-dissipative dynamics proposed in Ref.~\cite{qin2021generating} which stabilizes the spin cat states at long times, described in Sec.~\ref{sec:model}. The central theme of the paper is to address the following question: 
\begin{quote}\textit{In large ensembles of spins, can spin cat states with dissipative stabilization be robust against inhomogeneous broadening?}
\end{quote}

By analytically solving for the free evolution of the spin ensemble under the sole effect of the inhomogeneous broadening in Sec.~\ref{sec:parity}, we uncover an effect which we term \textit{parity-sensitive inhomogeneous dephasing}. We find that the robustness of the spin cat states to inhomogeneous broadening at small amplitudes depends critically on the parity symmetry of the state, such that the even cat state is significantly more robust to inhomogeneous dephasing than the odd cat state. At large amplitudes, the discrepancy between even and odd cat states vanishes, and the robustness of cat states are similar to that of a spin coherent state with the same amplitude.

Motivated by this observation, we focus on the robustness of the spin coherent state, in the presence of dissipative stabilization. To this end, we study in Sec.~\ref{sec:meanfield} the semiclassical mean-field dynamics derived from the full quantum model. The resulting dynamics can then be physically interpreted as a competition between spin synchronization and dephasing. We show that the system undergoes a synchronization phase transition, where the synchronization in the spin ensemble is completely broken in the long-time limit beyond a critical inhomogeneous broadening linewidth. This sets a limitation to the stability of the collective spin states against inhomogeneous dephasing. While the mean-field model is, by construction, inadequate in describing quantum effects, we argue that the analysis provides useful insights into the stability of spin cat states with large amplitudes in the full quantum model. Importantly, our findings suggest that in a realistic implementation of the quantum model, the spin cat state of either parity at large amplitudes is robust under the condition that the broadening linewidth does not scale with the system size. Our results provide a physical understanding of the robustness of spin cat states in realistic environments, which would be important in developing quantum technologies based on spin ensembles. We provide an outlook in Sec.~\ref{sec:outlook}. 

\section{Driven-dissipative spin model}
\label{sec:model}
We consider a system of $N$ two-level systems (TLS), i.e., pseudospin-$1/2$ particles, which we henceforth refer to as spins. The system is described by the Lindblad master equation
\begin{equation}
    \dot{\rho} = -i[H,\rho] + \frac{\Gamma_2}{N^2} \mathcal{D}\left[ \sum_{m,n=1}^{N} \sigma_m^- \sigma_n^- \right]\rho
\label{eq:ME}
\end{equation}
with the Hamiltonian in the rotating frame (setting $\hbar = 1$)
\begin{equation}
    H = \frac{1}{2} \sum_{n=1}^{N} \delta_n \sigma_n^z + \frac{\eta}{N} \left(\sum_{m,n = 1}^{N} e^{i\varphi} \sigma_m^- \sigma_n^- + e^{-i\varphi} \sigma_m^+ \sigma_n^+ \right).
\end{equation}
The parameters $\eta, \varphi$ and $\Gamma_2$ describe the squeezing strength, squeezing phase and nonlinear two-excitation loss rate respectively. $\sigma_n^{-} = \ket{g_n}\bra{e_n}$ is the lowering operator for the $n$-th particle from the excited state $\ket{e_n}$ to the ground state $\ket{g_n}$, and $\sigma_n^+ = \ket{e_i} \bra{g_n}$ is the corresponding raising operator. $\sigma_n^z = \ket{e_n}\bra{e_n} - \ket{g_n}\bra{g_n}$ Each spin has a detuning of $\delta_n$ compared to the ensemble mean frequency, which models the inhomogeneous broadening in the system. The dissipator $\mathcal{D}[A]\rho \equiv A \rho A^\dag - \{A^\dag A,\rho\}/2$ governs the dissipative interactions between the particles, with the jump operator $A$. The model in Eq.~\eqref{eq:ME} can be physically realized by collective two-photon coupling between the spins and a bosonic mode, such as an optical cavity, which is strongly dissipative. The bosonic mode can then be adiabatically eliminated, resulting in effective two-body dissipative interactions between the spins~\cite{qin2021generating}. A physical derivation of the master equation is provided in Appendix~\ref{app:derivationME}.

In the absence of inhomogeneous broadening (i.e., $\delta_n = 0$) and the regime $\eta \ll \Gamma_2$, the system is weakly excited and behaves as the bosonic model studied in~\cite{Mirrahimi2014dynamically}. This can be seen by performing the Holstein-Primakoff transformation $S_- = (N - a^\dag a)^{1/2} a$, where $S_\pm = \sum_n \sigma_n^{\pm}$ are the collective spin lowering and raising operators, while $a, a^\dag$ are the bosonic annihilation and creation operators respectively. The spins behave collectively when $\delta_n = 0$ due to permutation symmetry. In the weak excitation regime $\braket{a^\dag a} \ll N$, we have approximately $S_- \approx \sqrt{N} a$ and $S_+ \approx \sqrt{N} a^\dag$. The model can be approximately described (in the rotating frame) by
\begin{equation}
    \dot{\rho} \approx -i[H_{\text{bosonic}},\rho] + \Gamma_2 \mathcal{D}[a^2]\rho 
\end{equation}
where
\begin{equation}
    H_{\text{bosonic}} \approx \eta (e^{i \varphi} a^2 + e^{-i\varphi} a^{\dag 2}).
\end{equation}
The steady state of this bosonic model is spanned by the coherent states $\ket{\pm \alpha}$ with complex amplitude $\alpha = \sqrt{2\eta/\Gamma_2} \exp{(-i(\varphi/2 + \pi/4))}$~\cite{Mirrahimi2014dynamically,Leghtas2015confining}. By forming even and odd superpositions of $\ket{\pm \alpha}$, one obtains stabilized cat states which can be used to encode a logical qubit, where the errors can be corrected autonomously. These cat states are robust against single-photon loss, which is the dominant noise source in superconducting cavities. By mapping single-photon loss to either a logical bit or phase flip, this scheme generates a biased noise qubit, with the noise bias increasing with the amplitude $|\alpha|$ of the cat state~\cite{Lescanne2020exponential}. 

In Ref.~\cite{qin2021generating}, it was proposed to use collective spin systems in the weak excitation limit for a similar logical encoding, where the steady states are now spin coherent states~\cite{Radcliffe1971some} which can be superposed to obtain spin cat states analogous to the bosonic case. The main motivation behind this is to leverage the collective enhancement of the coherent spin interactions, such that the amplitude of the spin cat states scale as $\sim \sqrt{N}$ which translates to stronger protection against noise. It was argued that the spin cat states are robust to inhomogeneous broadening even though the permutation symmetry is broken. We now show that, for small amplitudes, this depends heavily on the parity symmetry of the cat state.
\begin{figure}
\centering
\subfloat{%
\includegraphics[width=0.95\linewidth]{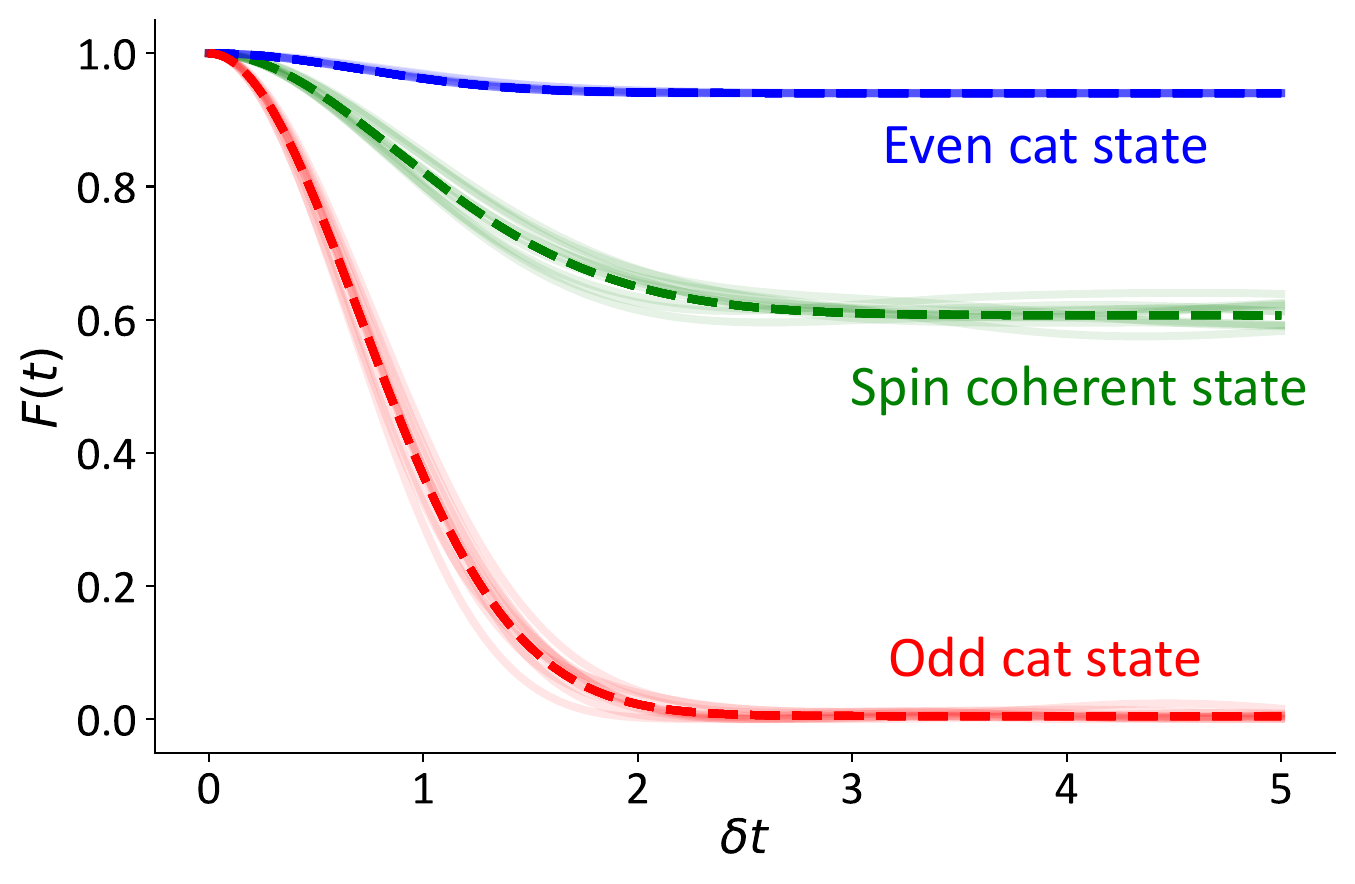}%
  \label{}%
}
\caption{Fidelity $F(t)$ for spin coherent, even cat and odd cat states under the free evolution of the dephasing Hamiltonian $H_0$. The light solid lines depict the random realizations of the detunings (10 realizations shown), while the dashed lines correspond to the analytical result for the mean fidelity $\overline{F(t)}$. The parameters are $N = 200$, $\theta = 1/\sqrt{N} \approx 0.0707, \phi = 0$.}
 \label{fig:fidelities}
\end{figure}
\section{Parity-sensitive inhomogeneous dephasing}
\label{sec:parity}
The spin coherent states are defined as
\begin{equation}
    \ket{\theta,\phi} = \bigotimes_{n=1}^{N} \left( \cos\frac{\theta}{2} \ket{g_n} + e^{i\phi} \sin \frac{\theta}{2} \ket{e_n} \right).
\label{eq:css}
\end{equation}

By considering even and odd superpositions of $\ket{\theta,\phi}$ and $\ket{\theta,\phi+\pi}$, one obtains the even and odd spin cat states
\begin{equation}
    \ket{\text{Cat}_{\pm}} = \frac{1}{\mathcal{N}_\pm} (\ket{\theta,\phi} \pm \ket{\theta,\phi+\pi}),
\end{equation}
with the normalization factors
\begin{equation}
    \mathcal{N}_\pm = \sqrt{2(1\pm \cos^N \theta)}.
\end{equation}
$\ket{\text{Cat}_\pm}$ are eigenstates of the parity operator
\begin{equation}
    \hat{\Pi} = \exp\left(i\pi\sum_n \sigma_n^+ \sigma_n^-\right)
\end{equation}
with eigenvalues $\pm 1$ respectively. Thus, $\ket{\text{Cat}_\pm}$ has even/odd number of excitations. Similar to the bosonic cat states, spin cat states are potentially useful for encoding a logical qubit and in quantum metrology. \dmtwo{For a sufficiently small inhomogeneous broadening, the dissipative model~\eqref{eq:ME} approximately encodes arbitrary superpositions of the spin coherent states in the steady state manifold. Thus, the spin cat states $\ket{\text{Cat}_\pm}$ can be prepared by initializing the spins in a state of definite parity (i.e., an eigenstate of $\hat{\Pi}$). For example, by initializing the spins in the ground state (corresponding to the vacuum state in the bosonic picture), the steady state is approximately $\ket{\text{Cat}_+}$. The validity of this approximation for the steady state depends on the stability of the cat states under inhomogeneous broadening, which is the core focus of this paper.}

In this section, we study the robustness of the spin cat states against inhomogeneous broadening. The main result here is that for small amplitudes $|\alpha| \ll 1$, the dephasing effect is sensitive to the parity of the cat state, with $\ket{\text{Cat}_+}$ being significantly more robust than $\ket{\text{Cat}_-}$. This discrepancy vanishes at large amplitudes $|\alpha| \gg 1$, and both $\ket{\text{Cat}_\pm}$ experience similar dephasing effects.

\subsection{Free evolution with inhomogeneous dephasing}
\label{sec:free_evol}

Let us first consider the dynamics of the spin ensemble under the free evolution governed by the dephasing (inhomogeneous broadening) Hamiltonian $H_0 = \sum_n \delta_n \sigma_n^z/2$. For concreteness, the detuning frequencies $\delta_n \sim \mathcal{N}(0,\delta^2)$ are independently and identically distributed random variables drawn from a Gaussian distribution with zero mean and variance $\delta^2$. The overlap between the evolved state and the initial spin coherent state is given by
\begin{equation}
\begin{split}
    c(t) &\equiv \braket{\theta,\phi|e^{-iH_0t}|\theta,\phi} \\&= \prod_{n=1}^{N} e^{i \delta_n t/2} \left( \cos^2 \frac{\theta}{2} + e^{-i\delta_n t} \sin^2 \frac{\theta}{2} \right).
\end{split}
\end{equation}
The fidelity $F(t) = |c(t)|^2$ indicates the survival probability of the initial state at time $t$, which evaluates to
\begin{equation}
\begin{split}
    F(t) &= \prod_{n=1}^{N} \left| \cos^2 \frac{\theta}{2} + e^{-i \delta_n t} \sin^{2} \frac{\theta}{2} \right|^2 \\&= \prod_{n=1}^{N} \left[ \cos^4 \frac{\theta}{2} + \sin^4 \frac{\theta}{2} + 2 \cos(\delta_n t) \sin^2 \frac{\theta}{2} \cos^2 \frac{\theta}{2} \right].
\end{split}
\end{equation}
Using the Gaussian average $\mathbb{E}[\cos \delta_n t] = e^{-\delta^2 t^2 / 2}$, we get the disorder-averaged fidelity
\begin{equation}
\begin{split}
    \overline{F(t)} &= \left[ \cos^4 \frac{\theta}{2} + \sin^4 \frac{\theta}{2} + 2 e^{-\delta^2 t^2/2} \sin^2 \frac{\theta}{2} \cos^2 \frac{\theta}{2} \right]^N \\&= \left[1 - 2\left(1-e^{-\delta^2 t^2/2}\right) \sin^2 \frac{\theta}{2} \cos^2 \frac{\theta}{2} \right]^N \\&= \left[ 1 - \frac{1}{2} \left(1-e^{-\delta^2 t^2 / 2}\right) \sin^2 \theta \right]^N,
\end{split}
\end{equation}
where the average is taken over random realizations of the frequencies $\delta_n$. In the weak excitation regime, $\theta$ is related to the amplitude of the bosonic coherent state via $|\alpha| \approx \sqrt{N} \tan(\theta/2)$~\cite{qin2021generating}. We will work in the regime where the bosonic amplitude $|\alpha| =\sqrt{2\eta/\Gamma_2} \ll \sqrt{N}$ such that $\theta \approx 2|\alpha|/\sqrt{N} \ll 1$. Note that $|\alpha|$ can still be much greater than $1$ for a sufficiently large $N$. Hence,
\begin{equation}
    \overline{F(t)} \approx \exp\left(-2|\alpha|^2(1-e^{-\delta^2 t^2/2}) \right).
\end{equation}
For small amplitudes $|\alpha| \ll 1$,
\begin{equation}
    \overline{F(t)} = 1 - 2|\alpha|^2 (1 - e^{-\delta^2 t^2 / 2}) + O(|\alpha|^4)
\end{equation}
which implies $\overline{F(\infty)} \approx 1 - 2|\alpha|^2$ is of order unity. Using $\mathbb{E}[\cos^2 \delta_n t] = (1+e^{-2\delta^2 t^2})/2$, we can also evaluate the variance of the fidelity,
\begin{equation}
    \text{Var}[F(t)] = \frac{2|\alpha|^4}{N} (1 - e^{-\delta^2 t^2}) + O(|\alpha|^6).
\label{eq:css_variance}
\end{equation}

Thus, the fluctuations of $F(t)$ are $\sim |\alpha|^2/\sqrt{N}$, much smaller than $\overline{F(t)}$. This implies that for sufficiently large systems, the random variable $F(t)$ concentrates around the mean $\overline{F(t)}$. 

As with the spin coherent states, we compute the fidelity of the cat states under the dephasing dynamics
\begin{equation}
    F_\pm(t) \equiv |\braket{\text{Cat}_\pm | e^{-iH_0 t}|\text{Cat}_\pm}|^2.
\label{eq:fid_pm}
\end{equation}
Using the Gaussian average $\mathbb{E}[\sin \delta_n t] = 0$ from symmetry, and after some algebra, we obtain
\begin{equation}
\begin{split}
    \overline{F_\pm (t)} &= \frac{4}{\mathcal{N}_\pm^4} \bigg[ \bigg(1-\frac{1}{2} (1-e^{-\delta^2 t^2/2}) \sin^2 \theta\bigg)^N \\&+ \bigg(1-\frac{1}{2} (1+e^{-\delta^2 t^2/2}) \sin^2 \theta\bigg)^N \pm 2 \cos^N \theta \bigg].
\end{split}
\end{equation}
Using the approximation
\begin{equation}
    \cos^N{\theta} \approx \left(1 - \frac{2|\alpha|^2}{N}\right)^{N} \approx e^{-2|\alpha|^2},    
\end{equation}
the fidelity is expressed as
\begin{equation}
\begin{split}
    \overline{F_\pm(t)} &\approx \frac{1}{(1 \pm e^{-2|\alpha|^2})^2}\bigg[  \exp\left(-2|\alpha|^2(1-e^{-\delta^2 t^2/2})\right) \\&+  \exp\left(-2|\alpha|^2(1+e^{-\delta^2 t^2/2})\right) \pm 2e^{-2|\alpha|^2} \bigg].
\end{split}
\label{eq:fid_averaged_alpha}
\end{equation}
In the long-time limit $t \to \infty$, we have
\begin{equation}
\begin{split}
    \overline{F_\pm(\infty)} &\approx \frac{2}{(1 \pm e^{-2|\alpha|^2})^2}\left(e^{-2|\alpha|^2} \pm e^{-2|\alpha|^2}\right) \\&= \begin{cases}
        \sech^2 (|\alpha|^2) \quad &(\text{even cat}), \\ 0 \quad &(\text{odd cat}).
    \end{cases}
\end{split}
\end{equation} 
This shows that $\ket{\text{Cat}_-}$ always decays to zero. On the other hand, the long-time fidelity for $\ket{\text{Cat}_+}$ depends on the amplitude. For small amplitude $|\alpha| \ll 1$, $\overline{F_+(\infty)} = 1 - |\alpha|^4 + O(|\alpha|^8)$ is robust, while for large amplitude $|\alpha| \gg 1$, $\overline{F_+(\infty)} \approx 4 e^{-2|\alpha|^2}$ vanishes exponentially with $|\alpha|$.

Next, we look at the short-time behavior of $\overline{F_\pm(t)}$. Expanding Eq.~\eqref{eq:fid_averaged_alpha} to order $t^2$ gives
\begin{equation}
    \overline{F_\pm(t)} \approx \begin{cases}
        1 - |\alpha|^2 \tanh(|\alpha|^2) \delta^2 t^2 \quad &\text{(even cat)}, \\ 1 - |\alpha|^2 \coth(|\alpha|^2) \delta^2 t^2 \quad &\text{(odd cat)}.
    \end{cases}
\label{eq:fid_avg_free}
\end{equation}
For small amplitude $|\alpha| \ll 1$,
\begin{equation}
    \overline{F_\pm(t)} \approx \begin{cases}
        1 - |\alpha|^4 \delta^2 t^2 \quad &\text{(even cat)}, \\ 1 - \delta^2 t^2 \quad &\text{(odd cat)},
    \end{cases}    
\end{equation}
implying that $\ket{\text{Cat}_-}$ decays faster than $\ket{\text{Cat}_+}$. For large amplitudes $|\alpha| \gg 1$, $\overline{F_\pm(t)} \approx 1 - |\alpha|^2 \delta^2 t^2$ and both $\ket{\text{Cat}_\pm}$ decay identically, and similarly as the spin coherent state. As with Eq.~\eqref{eq:css_variance}, one can also show that the relative fluctuations for the fidelity vanish as $N \to \infty$. 


This reveals an important fact that the robustness of the cat state against inhomogeneous broadening is very sensitive to its parity symmetry at small amplitudes. For the even cat state, the fidelity saturates at $F_+ \approx 1 - |\alpha|^2$ which is of order unity, while for the odd cat state the fidelity vanishes rapidly on the timescale of $\delta^{-1}$. Moreover, the even cat state is more robust than the spin coherent state, since the infidelity $1-\overline{F_+(t)} \sim |\alpha|^4$ (as compared to $|\alpha|^2$ for the spin coherent state). Physically, this means that the dephasing effects are suppressed by even symmetry and amplified by odd symmetry. The parity-sensitive dephasing of the cat states, at small amplitudes $|\alpha| \ll 1$, can be understood as an interference effect, where constructive and destructive interference occur for the even and odd cat states respectively. The strength of the interference is given by $|\braket{\theta,\phi|\theta,\phi+\pi}|^2 = \cos^{4N} \theta$, which appears in the cross terms of Eq.~\eqref{eq:fid_pm} upon expanding $\ket{\text{Cat}_\pm}$ in terms of the spin coherent states. At sufficiently large amplitudes, the interference strength is close to zero (as $N \to \infty)$, which is equivalent to the fact that $\ket{\theta,\phi}$ and $\ket{\theta,\phi+\pi}$ become nearly orthogonal. Consequently, the discrepancy in the fidelity decay between spin cat states of different parities vanishes at large amplitudes.

\subsection{Adding the stabilization dynamics}
\begin{figure}
\centering
\subfloat{%
\includegraphics[width=0.99\linewidth]{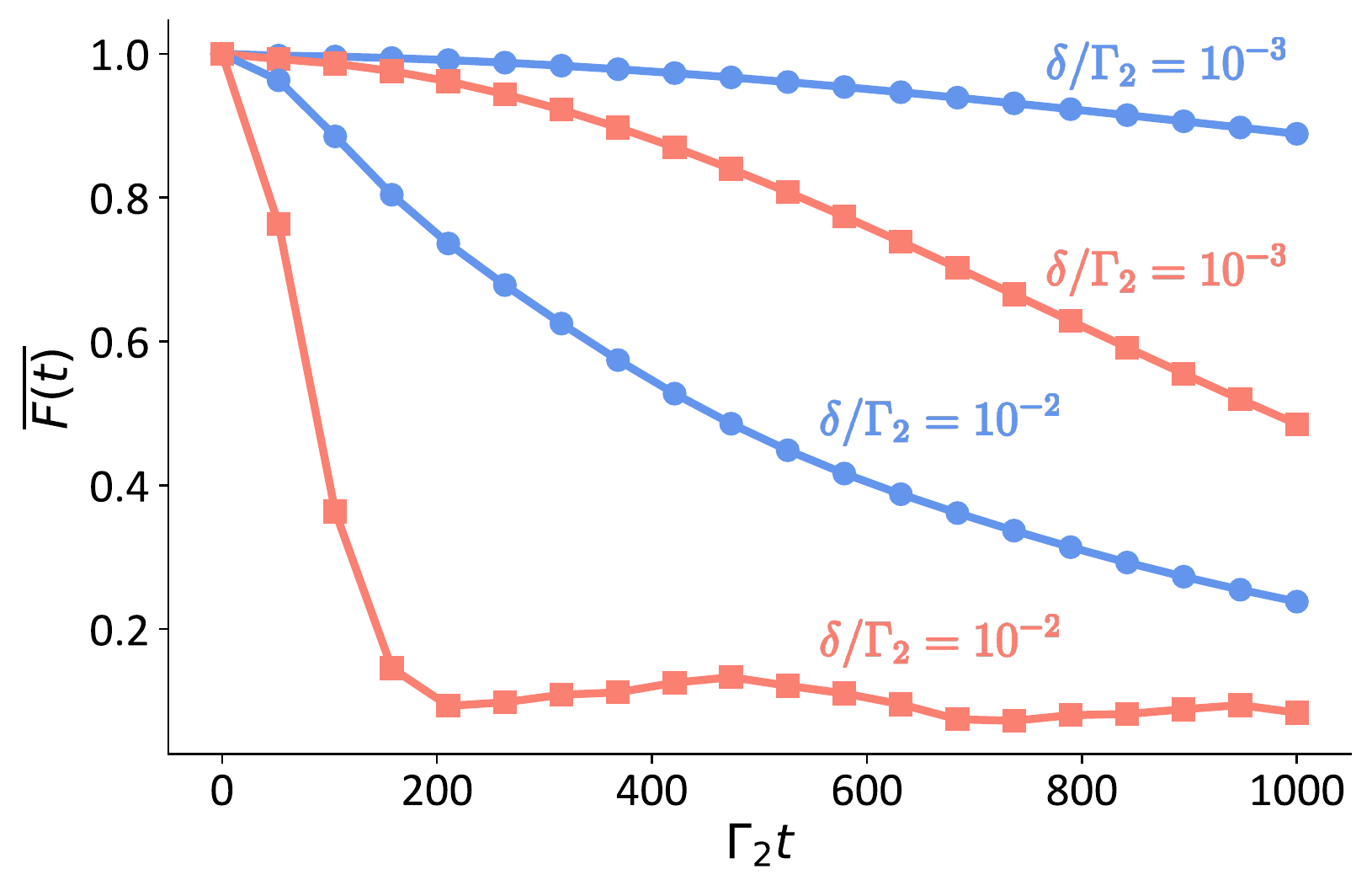}%
  \label{}%
}
\caption{Mean fidelity $\overline{F(t)}$ under the full dissipative dynamics~\eqref{eq:ME}. The $N = 8$ spins are initialized in the even (blue circles) and odd (red squares) cat states $\ket{\text{Cat}_\pm}$ with $\eta/\Gamma_2 = 0.2$, subject to inhomogeneous broadening with strengths $\delta/\Gamma_2 = \{10^{-3},10^{-2}\}$. The fidelities are computed with respect to the initial state and averaged over $10$ realizations of the detunings. In the absence of inhomogeneous broadening, the fidelities of the steady states are $0.998$ (even) and $0.995$ (odd).}
 \label{fig:MEfidelities}
\end{figure}
So far, we have only considered the free evolution governed by the dephasing Hamiltonian $H_0$. We now study numerically the full dissipative evolution governed by the master equation~\eqref{eq:ME}, using the QuTiP package~\cite{Johansson2013qutip}. Fig.~\ref{fig:MEfidelities} shows the mean fidelities $\overline{F(t)}$ for the cases of even and odd cat states, for $N = 8$. In each case, we initialize the system in the even/odd cat state and measure the fidelity with respect to the initial state. In the absence of inhomogeneous broadening, the steady states achieve fidelities of $0.998$ and $0.995$ for the even and odd cat states respectively. We see that the odd cat state is significantly more susceptible to the dephasing effects. For example, at time $\Gamma_2 t \approx 200$ with $\delta/\Gamma_2 = 10^{-2}$, the even cat state has a fidelity around $0.74$ while the fidelity of the odd cat state is only around $0.1$. This shows that the parity-sensitive dephasing derived for the simple case of only free Hamiltonian evolution applies even with the driven-dissipative stabilization terms. In Ref.~\cite{qin2021generating}, it was shown numerically that the even cat state $\ket{\text{Cat}_+}$ is robust against inhomogeneous broadening (and other imperfections). This is consistent with our findings. We argue here that the same robustness does not apply to the odd cat state, even at larger system sizes beyond the reach of numerical simulations. For larger $N$, one expects that the effects of inhomogeneous broadening are suppressed due to stronger collective interactions, but the separation of timescales in the fidelity decay between the even and odd cat state should remain valid, at small amplitudes. 

\begin{figure}
\centering
\subfloat{%
\includegraphics[width=0.99\linewidth]{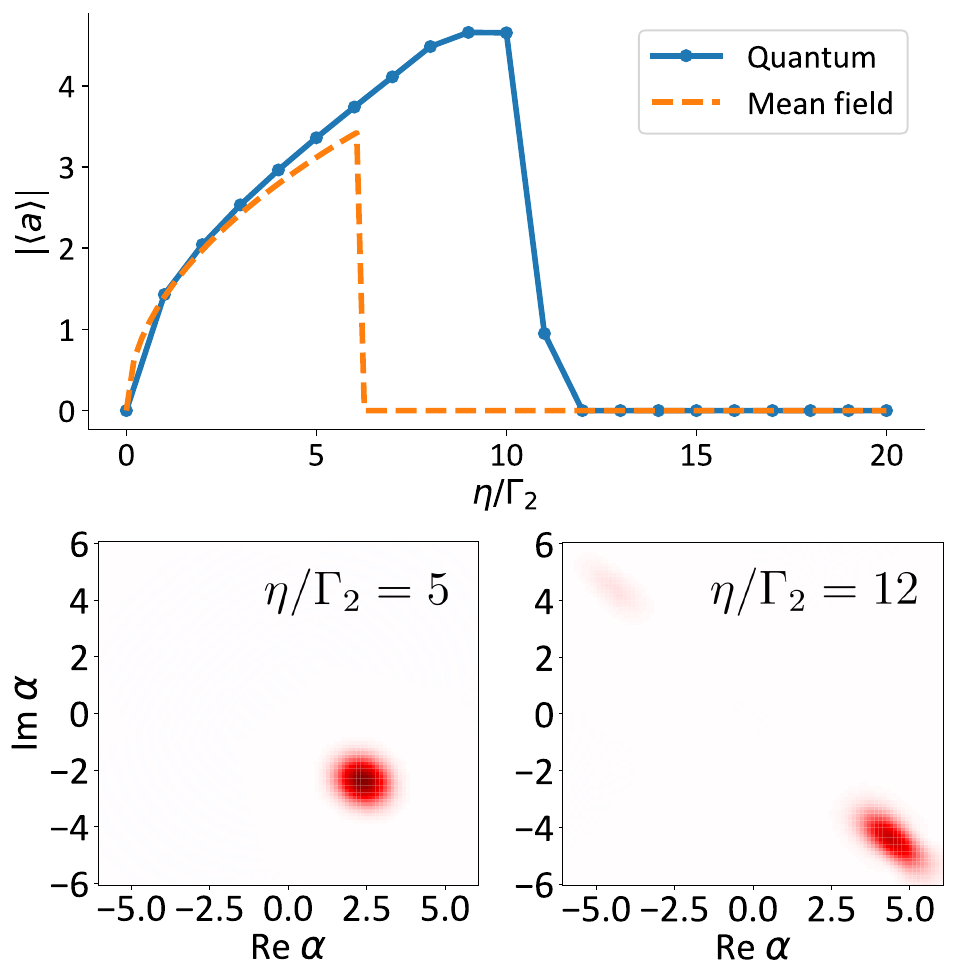}%
  \label{}%
}
\caption{(Top) Steady state amplitude $|\braket{a}|$ against $\eta/\Gamma_2$ for the quantum model with $N = 100$. The initial state is the bosonic coherent state with complex amplitude $\alpha = \sqrt{2\eta/\Gamma_2} e^{-i\pi/4}$. The dashed line corresponds to the mean field solution for $\sqrt{N} A$ in Eq.~\eqref{eq:oneensemble_ss}. (Bottom) Steady state Wigner functions for $\eta/\Gamma_2 \in \{5,12\}$.}
 \label{fig:amplitudes}
\end{figure}

\section{Mean-field analysis: Synchronization phase transition}
\label{sec:meanfield}
Since a full analytical treatment of Eq.~\eqref{eq:ME} is not feasible, we consider instead a mean-field approximation where we assume a product state ansatz for $\rho$. The spin coherent state in Eq.~\eqref{eq:css} is a product state, so we expect this to be a reasonable approximation for large $N$. However, note that the mean-field approximation cannot capture superpositions and thus does not account for parity-sensitive dephasing. Nonetheless, for large amplitudes, the discrepancy in the dephasing effects between $\ket{\text{Cat}_\pm}$ becomes irrelevant, and behaves similarly as the spin coherent state. Thus, we expect the study of the mean-field model to yield useful insights about the stability of the spin cat states in the full quantum model~\eqref{eq:ME}.

The mean-field equations can be physically interpreted as describing the synchronization dynamics of $N$ globally coupled classical spins, containing both dissipative and coherent (also called reactive) couplings. As we will show, this system exhibits a synchronization transition with both $\eta/\Gamma_2$ and $\delta/\Gamma_2$, where the spins become desynchronized beyond a critical parameter value. This provides important insights on the robustness of the spin coherent states to inhomogeneous broadening with driven-dissipative stabilization, for large system sizes.   

The mean-field equations governing the dynamics of the $m$-th spin read (see Appendix~\ref{app:meanfield} for a detailed derivation):
\begin{equation}
\begin{split}
    \frac{d}{dt}\braket{\sigma_m^+} &= i \delta_m \braket{\sigma_m^+} - 2i\eta e^{i\theta} \braket{\sigma_m^z} c_{1,m}^* \\&- 2 \Gamma_2 \braket{\sigma_m^+} \left(\frac{c_{2,m}}{N} + |c_{1,m}|^2 \right) \\&+ \Gamma_2 \braket{\sigma_m^z} c_{1,m} (2c_{2,m} + N |c_{1,m}|^2 )
\end{split}
\label{eq:mf_sigma+}
\end{equation}
and
\begin{equation}
\begin{split}
    \frac{d}{dt}\braket{\sigma_m^z} &= 8\eta \text{Im} \left( e^{-i\theta} \braket{\sigma_m^+}c_{1,m} \right) \\&- 4\Gamma_2 (1+\braket{\sigma_m^z}) \left(\frac{c_{2,m}}{N} + |c_{1,m}|^2 \right) \\&- 4\Gamma_2 \text{Re} [\braket{\sigma_m^+} c_{1,m}^* (2c_2 + N|c_{1,m}|^2 ) ]
\end{split}
\label{eq:mf_sigmaz}
\end{equation}
where $c_{1,m} \equiv \frac{1}{N} \sum_{j \neq m} \braket{\sigma_j^+}$ is the average coherence and $c_{2,m} \equiv \frac{1}{2N} \sum_{j \neq m} (1 + \braket{\sigma_j^z})$ is the spin excitation density, excluding the $m$-th spin. \text{Re} and \text{Im} denote the real and imaginary parts respectively. The squeezing phase $\varphi$ is set to zero without loss of generality.

\subsection{Identical frequencies}
First, we consider the case where all the spins have the same frequency ($\delta = 0$). This recovers the permutation symmetry in the system, which allows us to compare the mean-field solution to the exact quantum dynamics. From symmetry arguments, we denote $\braket{\sigma_j^+} = \braket{\sigma^+} = A \exp(i \phi)$ where $A$ and $\phi$ are to be determined, and $\braket{\sigma_j^z} = \braket{\sigma^z} = z$. Then, $c_{1,m} = (N-1)/N \times A \exp(i\phi)$ and $c_{2,m} = (N-1)/2N \times (1 + z)$. This reduces the problem drastically from $3N$ real variables to just $3$ real variables $A$, $\phi$ and $z$. For large $N$, the reduced equations of motion become
\begin{equation}
\begin{split}
    \frac{\dot{A}}{A} &= -2\eta z \sin 2\phi - 2 \Gamma_2 A^2 + N \Gamma_2 z A^2 \\
    \dot{\phi} &= -2\eta z A \cos 2\phi \\
    \frac{\dot{z}}{A^2} &= 8\eta \sin 2\phi - 4 \Gamma_2 (1 + z) - 4 \Gamma_2 N A^2. 
\end{split}
\label{eq:oneensemble_MF}
\end{equation}
In the low excitation limit, the steady state of the master equation~\eqref{eq:ME} is approximately the bosonic coherent state with $|\braket{a}| = \sqrt{N} A = \sqrt{2\eta/\Gamma_2}, z \approx -1, \phi = \pi/4$. This agrees with the steady state solution of Eq. \eqref{eq:oneensemble_MF}:
\begin{equation}
\begin{split}
    A^2 &\approx \frac{4\eta - 1 + \sqrt{1- \frac{16\eta}{N\Gamma_2}}}{2N} \\
    \phi &= \pi/4 \\
    z &\approx -\frac{1+\sqrt{1-\frac{16\eta}{N\Gamma_2}}}{2}
\end{split}
\label{eq:oneensemble_ss}
\end{equation}
which is valid in the range $0 \leq \eta/(N\Gamma_2) \leq 1/16$. To leading order in $\eta/{N \Gamma_2}$, we recover $A^2 = 2\eta/N\Gamma_2$ and $z = -1$ which matches the quantum results in the low excitation limit. The solution in Eq. \eqref{eq:oneensemble_ss} describes the synchronized state where the spins are all phase locked to one another. The range of validity of this solution can be interpreted as the synchronization region, where the synchronization is broken at $\eta/(N\Gamma_2) = 1/16$. This sharp behavior in $\eta/\Gamma$ can also be understood for the quantum system. Making the exact Holstein-Primakoff transformation as introduced in Sec.~\ref{sec:model}, we plot the steady state bosonic amplitude $|\braket{a}|$ in Fig.~\ref{fig:amplitudes} for $N = 100$ spins, initializing in the coherent state with complex amplitude $\alpha = \sqrt{2\eta/\Gamma_2} e^{-i\pi/4}$. For small $\eta/\Gamma_2$, the mean-field approximation agrees well with the quantum model. Interestingly, the bosonic amplitude increases and drops sharply at $\eta/\Gamma_2 \approx 10$, although the drop occurs at a larger value of $\eta/\Gamma_2$ compared to the mean field model ($\eta/\Gamma = 6.25$). In the quantum case, the drop in amplitude has a different physical interpretation: the bosonic Hilbert space has a dimension of $N + 1$. Due to the boundedness of the quantum phase space, the bosonic amplitude cannot increase indefinitely with $\eta/\Gamma_2$ as $\sqrt{2\eta/\Gamma_2}$. Thus, for sufficiently large amplitudes, a secondary blob emerges in the Wigner function at a phase difference of $\pi$ and causing $|\braket{a}|$ to collapse. As a simple estimate, this collapse occurs when $\braket{a^\dag a} \propto N$ such that the boundaries of the phase space become relevant. Since $\braket{a^\dag a} \sim \eta/\Gamma_2$, the amplitude collapse occurs when $\eta/\Gamma_2 \propto N$, the same scaling as the mean-field prediction (even though the exact values differ).

\subsection{Oppositely detuned sub-ensembles}
Now, we break permutation symmetry by adding detunings to the spins. For analytical tractability, we consider two sub-ensembles, each with $N/2$ spins (assuming $N$ is even), with detunings $\pm \delta$. The results derived for this model should hold qualitatively for the more realistic model of random detunings (e.g., Gaussian distributed) which describes inhomogeneous broadening. Assuming that $\braket{\sigma^+} = A \exp[i(\pi/4 \pm \zeta)]$ for the two sub-ensembles respectively, where $\zeta$ is the deviation of the phase from $\pi/4$, and $\braket{\sigma_j^z} = z$ for all spins, we can reduce the problem once again to just 3 nonlinear coupled equations
\begin{equation}
\begin{split}
    \frac{\dot{A}}{A} &= -2\eta z \left(\cos^2 \zeta - \frac{1}{N} \cos 2\zeta\right) \\&+ \Gamma_2 z \left(\cos^2 \zeta - \frac{1}{N} \right) \left[ 1+z + NA^2 \left(1-\frac{2}{N}\right) \cos^2 \zeta \right] \\&- 2 \Gamma_2 A^2 \cos^2 \zeta, 
\end{split}
\label{eq:twoensemble_MF1}
\end{equation}

\begin{equation}
\begin{split}
    \dot{\zeta} &= \delta + \eta z \left(1-\frac{2}{N}\right) \sin 2\zeta \\&- \frac{\Gamma_2}{2} z \left[ 1+z + NA^2 \left(1-\frac{2}{N}\right) \cos^2 \zeta \right] \sin 2\zeta,
\end{split}
\label{eq:twoensemble_MF2}
\end{equation}
and
\begin{equation}
\begin{split}
\frac{\dot{z}}{A^2} &= 8\eta \left(\cos^2 \zeta - \frac{1}{N} \cos 2\zeta \right) \\&- 4\Gamma_2 \left(\cos^2 \zeta - \frac{1}{N}\right) \left[ 1+z + NA^2 \left(1-\frac{2}{N}\right) \cos^2 \zeta \right] \\&- 4\Gamma_2 (1+z) \cos^2 \zeta.
\end{split}
\label{eq:twoensemble_MF3}
\end{equation}

To verify the accuracy of Eqs.~(\ref{eq:twoensemble_MF1}),\,(\ref{eq:twoensemble_MF2}), and (\ref{eq:twoensemble_MF3}), we numerically simulate them and compare the results with the full mean-field equations~\eqref{eq:mf_sigma+} and~\eqref{eq:mf_sigmaz}. It is also simple to see that this is consistent with Eq.~\eqref{eq:oneensemble_ss} by setting $\delta = 0, \phi \approx \pi/4, z \approx -1$ and taking the large $N$ limit.

Compared to the previous case of identical frequencies, this set of equations is much harder to solve analytically. For small $\zeta$, we can get an approximate solution by expanding Eqs.~(\ref{eq:twoensemble_MF1}),\,(\ref{eq:twoensemble_MF2}), and (\ref{eq:twoensemble_MF3}) to first order in $\zeta$ and also first order in $1/N$. The approximate steady state solution corresponding to the synchronized state is (see Appendix \ref{app:twoensembles})
\begin{equation}
\begin{split}
    A^2 &\approx \frac{4\eta - 1 + \sqrt{1- \frac{16\eta}{N\Gamma_2}}}{2N} \\
    \zeta &\approx \frac{1+\sqrt{1-\frac{16\eta}{N\Gamma_2}}}{32 \eta^2} N^2 \Gamma_2 \delta \\
    z &\approx -\frac{1+\sqrt{1- \frac{16\eta}{N\Gamma_2}}}{2}
\end{split}
\label{eq:twoensemble_ss}
\end{equation}
This solution is valid for small $\delta$ (where the dephasing of the ensemble is slow) or large $\eta$ (where the synchronization effect is strong). For $\eta/N\Gamma_2 \ll 1$, the formula for phase spread simplifies to $\zeta \approx N^2 \Gamma_2 \delta / 16\eta^2$. Interestingly, while the phase spread is linear in $\delta$, the steady state values for $A^2$ and $z$ are independent of $\delta$ up to linear order. 
\begin{figure}
\centering
\subfloat{%
\includegraphics[width=0.99\linewidth]{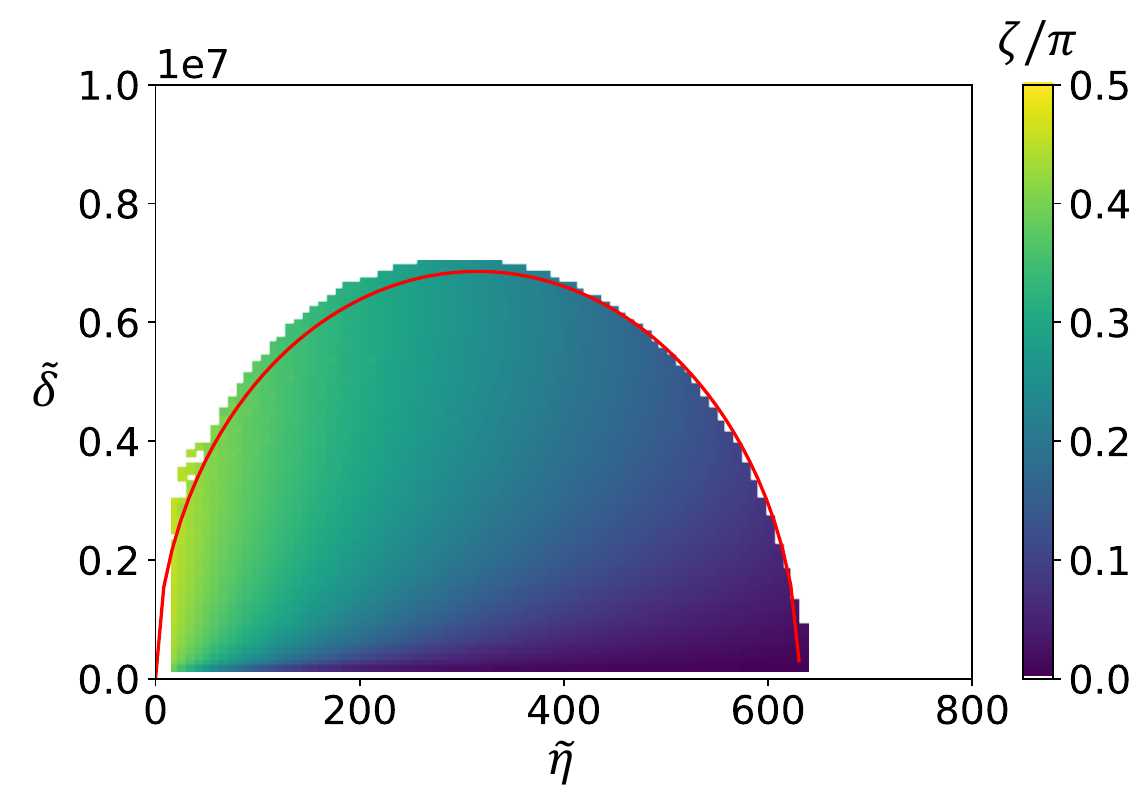}%
}
\caption{Steady state value of the phase spread $\zeta$ against the dimensionless detuning $\tilde{\delta} = N^2 \delta/\Gamma_2$ and dimensionless squeezing $\tilde{\eta} = \eta/\Gamma_2$, for $N = 10^4$ spins. The white region denote data points where the spins fail to converge to a synchronized state at long times. The red line marks the elliptical boundary of the synchronization region given by Eq.~\eqref{eq:ellipse}, with fitting parameters $a \approx 0.03125N$ and $b \approx 2.2077N$. For small values of $\tilde{\eta}$, some points within the synchronization region fail to converge to the steady state within the simulation time due to the slow dynamics.}
 \label{fig:synctransition}
\end{figure}
For larger values of $\delta$, we perform numerical simulations of the mean-field equations for $N=10^4$ spins, shown in Fig.~\ref{fig:synctransition}. The steady state value of the phase spread $\zeta$ is small for small $\delta$ and large $\eta$, which agrees with our physical intuition. For the case of identical frequencies $\delta = 0$, there is a maximum $\eta / \Gamma_2$ of $N/16 = 625$ beyond which the synchronization breaks, agreeing with our analytical calculations. As $\delta$ increases, the region of synchronization narrows until a threshold value such that no synchronization is possible for any $\eta$. The boundary of the synchronization region appears to be elliptical. We postulate that the boundary curve takes the form
\begin{equation}
    \tilde{\delta} = b \sqrt{a^2 - (\tilde{\eta} -a )^2}
\label{eq:ellipse}
\end{equation}
where $\tilde{\delta} \equiv N^2 \delta / \Gamma_2$ and $\tilde{\eta} \equiv \eta / \Gamma_2$ are the dimensionless frequency detuning and squeezing respectively, and $a,b$ are the fitting parameters. The threshold values are thus $\tilde{\eta}_c = 2a$ and $\tilde{\delta}_c = ab$. From numerical fitting, we find that $a \approx 0.03125N$ and $ b \approx 2.2077 N$ which gives $\tilde{\eta}_c \approx 0.0625N$ and $\tilde{\delta}_c \approx 0.06888N^2$. The value of $\tilde{\eta}_c$ agrees excellently with the theoretical prediction of $N/16$.


The mean-field analysis suggests that the spin coherent state is robust against inhomogeneous broadening, under the condition that the broadening linewidth does not scale with $N$. From Eq.~\eqref{eq:twoensemble_ss}, we have $\zeta \approx N^2 \delta / 4|\alpha|^4\Gamma_2$. This implies that $\delta/\Gamma_2$ needs to scale as $\sim 1/N^2$ for stability at a constant amplitude $|\alpha|$. This condition is satisfied for constant $\delta$ since $\Gamma_2 \sim N^2$ (see Appendix~\ref{app:derivationME}). A full quantum analysis of the robustness remains an open problem and is left as future work.

\section{Discussion and Outlook}
\label{sec:outlook}
In this work, we analyze the effects of symmetry-breaking inhomogeneous dephasing on collective spin states such as spin coherent states and cat states $|\text{Cat}_\pm\rangle$ formed from the superposition of two spin coherent states offset by a $\pi$-phase.

To summarize the core arguments of our paper:
\begin{enumerate}
    \item For cat states with small amplitudes $(|\alpha| \ll 1)$, the even spin cat state is more robust to inhomogeneous dephasing than the spin coherent state of the same amplitude, which is in turn more robust than the odd spin cat state. We demonstrate this parity-sensitive dephasing analytically without dissipative stabilization, which is supported with exact numerics of the Lindblad master equation for small system sizes.
    \item However, for large cat states $|\alpha| \gg 1$, the sensitivity to the number parity vanishes. Consequently, the robustness of the spin cat states is similar to that of a spin coherent state with the same $|\alpha|$. Thus, we can estimate the stability of large spin cat states by analyzing the spin coherent state.
    \item The stability of the spin coherent state (with dissipative stabilization) can be analyzed using a mean-field approach to the Lindblad master equation. The quantum model is now approximated by a synchronization model of classical spins.
    \item We then show analytically that the mean-field model exhibits a synchronization phase transition at a critical inhomogeneous broadening linewidth, beyond which the synchronization of the classical spins breaks down. We argue that this characterizes the robustness of the spin coherent states (and from point 2, the large cat states of any parity as well).
    \item By analyzing the robustness of the classical synchronization model, we estimate that the dissipatively stabilized spin cats (with amplitude $|\alpha| \gg 1$) are robust to inhomogeneous broadening if $\delta/\Gamma_2 \sim 1$. This is satisfied if the broadening linewidth $\delta$ does not increase proportionally to $N$, since $\Gamma_2 \sim N$ in a realistic implementation of the model.
\end{enumerate}


Our findings have important implications in using macroscopic spin ensembles as `cat qubits' to encode logical quantum information, motivated by recent successful experiments in superconducting quantum circuits. Due to the collectively enhanced interactions in spin ensembles scaling as $\sqrt{N}$, they stand to benefit from greater error-correcting capabilities compared to their bosonic counterpart. However, using spin ensembles also come with a different set of experimental challenges like particle loss which has to be carefully addressed. In certain implementations such as rare-earth ions, inhomogeneous broadening arising from spatially-varying magnetic fields break permutation symmetry which is required for the spins to behave collectively, and causes spin dephasing. In Ref.~\cite{qin2021generating}, it was demonstrated explicitly that $|\text{Cat}_+\rangle$ is robust against such inhomogeneous dephasing, which is consistent with our results. However, our results also suggest that the odd cat state $|\text{Cat}_-\rangle$ is significantly more fragile. This means that the lifetime of $|\text{Cat}_-\rangle$ is limited by the dephasing timescale from the inhomogeneous broadening, which is on the order of $1$ $\mu\text{s}$ (taking $\delta \sim 1$ MHz). Consequently, encoding the logical qubit using $|\text{Cat}_\pm\rangle$ is unlikely to result in logical lifetimes significantly longer than the break-even point, unless the bosonic amplitude is large, in which our results indicate that $\ket{\text{Cat}_-}$ can be potentially robust.

An alternative logical encoding is to define the basis states $\ket{0}_L \propto \ket{\theta,0} + \ket{\theta,\pi}$ and $\ket{1}_L \propto \ket{\theta,\pi/2} + \ket{\theta,3\pi/2}$, where $\ket{\theta,\phi}$ are the spin coherent states~\eqref{eq:css}. Here, the logical qubit is encoded in the even parity subspace which enjoys the enhanced protection against inhomogeneous broadening. This encoding is analogous to the `four-legged cat code' proposed in bosonic quantum error correction~\cite{Leghtas2013hardware}. To stabilize this as the steady state, one approach would be to replace the collective two-body terms in Eq.~\eqref{eq:ME} with collective four-body terms. However, it is very challenging in practice to implement quartic dissipators of the form $\mathcal{D}[S_-^4]\rho$, where $S_-$ is the collective spin lowering operator, while suppressing all unwanted dissipative terms.

Apart from quantum error correction, our results are also relevant for other applications of collective spin systems such as metrology and quantum simulation of dissipative phase transitions. As a future work, it would be interesting to study the parity-sensitive dephasing effect in more general dynamics beyond the specific master equation studied here, and also for general spin states with well-defined parity symmetry. One can also go beyond the product state ansatz and include quantum correlations in the numerical simulations using higher-order mean-field methods such as cumulant expansion~\cite{kubo1962generalized,Plankensteiner2022quantumcumulantsjl}, which has been widely employed to study many-body spin dynamics~\cite{Hotter2023cavity,debnath2018lasing,debnath2019collective,debnath2020self,rubies2023characterizing,masson2024dicke}.

\section*{Acknowledgments}
\dmtwo{We thank the anonymous referees for helpful suggestions.}
L.C.K. acknowledges support from the Ministry of Education, Singapore and the National Research Foundation,
Singapore. S.T acknowledges support from the National Research Foundation, Singapore. The Institute for Quantum Information and Matter is an NSF Physics Frontiers Center.

\bibliography{bib}

\begin{thebibliography}{39}%
\makeatletter
\providecommand \@ifxundefined [1]{%
 \@ifx{#1\undefined}
}%
\providecommand \@ifnum [1]{%
 \ifnum #1\expandafter \@firstoftwo
 \else \expandafter \@secondoftwo
 \fi
}%
\providecommand \@ifx [1]{%
 \ifx #1\expandafter \@firstoftwo
 \else \expandafter \@secondoftwo
 \fi
}%
\providecommand \natexlab [1]{#1}%
\providecommand \enquote  [1]{``#1''}%
\providecommand \bibnamefont  [1]{#1}%
\providecommand \bibfnamefont [1]{#1}%
\providecommand \citenamefont [1]{#1}%
\providecommand \href@noop [0]{\@secondoftwo}%
\providecommand \href [0]{\begingroup \@sanitize@url \@href}%
\providecommand \@href[1]{\@@startlink{#1}\@@href}%
\providecommand \@@href[1]{\endgroup#1\@@endlink}%
\providecommand \@sanitize@url [0]{\catcode `\\12\catcode `\$12\catcode
  `\&12\catcode `\#12\catcode `\^12\catcode `\_12\catcode `\%12\relax}%
\providecommand \@@startlink[1]{}%
\providecommand \@@endlink[0]{}%
\providecommand \url  [0]{\begingroup\@sanitize@url \@url }%
\providecommand \@url [1]{\endgroup\@href {#1}{\urlprefix }}%
\providecommand \urlprefix  [0]{URL }%
\providecommand \Eprint [0]{\href }%
\providecommand \doibase [0]{https://doi.org/}%
\providecommand \selectlanguage [0]{\@gobble}%
\providecommand \bibinfo  [0]{\@secondoftwo}%
\providecommand \bibfield  [0]{\@secondoftwo}%
\providecommand \translation [1]{[#1]}%
\providecommand \BibitemOpen [0]{}%
\providecommand \bibitemStop [0]{}%
\providecommand \bibitemNoStop [0]{.\EOS\space}%
\providecommand \EOS [0]{\spacefactor3000\relax}%
\providecommand \BibitemShut  [1]{\csname bibitem#1\endcsname}%
\let\auto@bib@innerbib\@empty
\bibitem [{\citenamefont {Pezz\`e}\ \emph {et~al.}(2018)\citenamefont
  {Pezz\`e}, \citenamefont {Smerzi}, \citenamefont {Oberthaler}, \citenamefont
  {Schmied},\ and\ \citenamefont {Treutlein}}]{Pezze2018Quantum}%
  \BibitemOpen
  \bibfield  {author} {\bibinfo {author} {\bibfnamefont {L.}~\bibnamefont
  {Pezz\`e}}, \bibinfo {author} {\bibfnamefont {A.}~\bibnamefont {Smerzi}},
  \bibinfo {author} {\bibfnamefont {M.~K.}\ \bibnamefont {Oberthaler}},
  \bibinfo {author} {\bibfnamefont {R.}~\bibnamefont {Schmied}},\ and\ \bibinfo
  {author} {\bibfnamefont {P.}~\bibnamefont {Treutlein}},\ }\bibfield  {title}
  {\bibinfo {title} {Quantum metrology with nonclassical states of atomic
  ensembles},\ }\href {https://doi.org/10.1103/RevModPhys.90.035005} {\bibfield
   {journal} {\bibinfo  {journal} {Rev. Mod. Phys.}\ }\textbf {\bibinfo
  {volume} {90}},\ \bibinfo {pages} {035005} (\bibinfo {year}
  {2018})}\BibitemShut {NoStop}%
\bibitem [{\citenamefont {Zhou}\ \emph {et~al.}(2020)\citenamefont {Zhou},
  \citenamefont {Choi}, \citenamefont {Choi}, \citenamefont {Landig},
  \citenamefont {Douglas}, \citenamefont {Isoya}, \citenamefont {Jelezko},
  \citenamefont {Onoda}, \citenamefont {Sumiya}, \citenamefont {Cappellaro},
  \citenamefont {Knowles}, \citenamefont {Park},\ and\ \citenamefont
  {Lukin}}]{Zhou2020quantum}%
  \BibitemOpen
  \bibfield  {author} {\bibinfo {author} {\bibfnamefont {H.}~\bibnamefont
  {Zhou}}, \bibinfo {author} {\bibfnamefont {J.}~\bibnamefont {Choi}}, \bibinfo
  {author} {\bibfnamefont {S.}~\bibnamefont {Choi}}, \bibinfo {author}
  {\bibfnamefont {R.}~\bibnamefont {Landig}}, \bibinfo {author} {\bibfnamefont
  {A.~M.}\ \bibnamefont {Douglas}}, \bibinfo {author} {\bibfnamefont
  {J.}~\bibnamefont {Isoya}}, \bibinfo {author} {\bibfnamefont
  {F.}~\bibnamefont {Jelezko}}, \bibinfo {author} {\bibfnamefont
  {S.}~\bibnamefont {Onoda}}, \bibinfo {author} {\bibfnamefont
  {H.}~\bibnamefont {Sumiya}}, \bibinfo {author} {\bibfnamefont
  {P.}~\bibnamefont {Cappellaro}}, \bibinfo {author} {\bibfnamefont {H.~S.}\
  \bibnamefont {Knowles}}, \bibinfo {author} {\bibfnamefont {H.}~\bibnamefont
  {Park}},\ and\ \bibinfo {author} {\bibfnamefont {M.~D.}\ \bibnamefont
  {Lukin}},\ }\bibfield  {title} {\bibinfo {title} {Quantum metrology with
  strongly interacting spin systems},\ }\href
  {https://doi.org/10.1103/PhysRevX.10.031003} {\bibfield  {journal} {\bibinfo
  {journal} {Phys. Rev. X}\ }\textbf {\bibinfo {volume} {10}},\ \bibinfo
  {pages} {031003} (\bibinfo {year} {2020})}\BibitemShut {NoStop}%
\bibitem [{\citenamefont {Arunkumar}\ \emph {et~al.}(2023)\citenamefont
  {Arunkumar}, \citenamefont {Olsson}, \citenamefont {Oon}, \citenamefont
  {Hart}, \citenamefont {Bucher}, \citenamefont {Glenn}, \citenamefont {Lukin},
  \citenamefont {Park}, \citenamefont {Ham},\ and\ \citenamefont
  {Walsworth}}]{Arunkumar2023quantum}%
  \BibitemOpen
  \bibfield  {author} {\bibinfo {author} {\bibfnamefont {N.}~\bibnamefont
  {Arunkumar}}, \bibinfo {author} {\bibfnamefont {K.~S.}\ \bibnamefont
  {Olsson}}, \bibinfo {author} {\bibfnamefont {J.~T.}\ \bibnamefont {Oon}},
  \bibinfo {author} {\bibfnamefont {C.~A.}\ \bibnamefont {Hart}}, \bibinfo
  {author} {\bibfnamefont {D.~B.}\ \bibnamefont {Bucher}}, \bibinfo {author}
  {\bibfnamefont {D.~R.}\ \bibnamefont {Glenn}}, \bibinfo {author}
  {\bibfnamefont {M.~D.}\ \bibnamefont {Lukin}}, \bibinfo {author}
  {\bibfnamefont {H.}~\bibnamefont {Park}}, \bibinfo {author} {\bibfnamefont
  {D.}~\bibnamefont {Ham}},\ and\ \bibinfo {author} {\bibfnamefont {R.~L.}\
  \bibnamefont {Walsworth}},\ }\bibfield  {title} {\bibinfo {title} {Quantum
  logic enhanced sensing in solid-state spin ensembles},\ }\href
  {https://doi.org/10.1103/PhysRevLett.131.100801} {\bibfield  {journal}
  {\bibinfo  {journal} {Phys. Rev. Lett.}\ }\textbf {\bibinfo {volume} {131}},\
  \bibinfo {pages} {100801} (\bibinfo {year} {2023})}\BibitemShut {NoStop}%
\bibitem [{\citenamefont {Wesenberg}\ \emph {et~al.}(2009)\citenamefont
  {Wesenberg}, \citenamefont {Ardavan}, \citenamefont {Briggs}, \citenamefont
  {Morton}, \citenamefont {Schoelkopf}, \citenamefont {Schuster},\ and\
  \citenamefont {M\o{}lmer}}]{Wesenberg2009quantum}%
  \BibitemOpen
  \bibfield  {author} {\bibinfo {author} {\bibfnamefont {J.~H.}\ \bibnamefont
  {Wesenberg}}, \bibinfo {author} {\bibfnamefont {A.}~\bibnamefont {Ardavan}},
  \bibinfo {author} {\bibfnamefont {G.~A.~D.}\ \bibnamefont {Briggs}}, \bibinfo
  {author} {\bibfnamefont {J.~J.~L.}\ \bibnamefont {Morton}}, \bibinfo {author}
  {\bibfnamefont {R.~J.}\ \bibnamefont {Schoelkopf}}, \bibinfo {author}
  {\bibfnamefont {D.~I.}\ \bibnamefont {Schuster}},\ and\ \bibinfo {author}
  {\bibfnamefont {K.}~\bibnamefont {M\o{}lmer}},\ }\bibfield  {title} {\bibinfo
  {title} {Quantum computing with an electron spin ensemble},\ }\href
  {https://doi.org/10.1103/PhysRevLett.103.070502} {\bibfield  {journal}
  {\bibinfo  {journal} {Phys. Rev. Lett.}\ }\textbf {\bibinfo {volume} {103}},\
  \bibinfo {pages} {070502} (\bibinfo {year} {2009})}\BibitemShut {NoStop}%
\bibitem [{\citenamefont {Brion}\ \emph {et~al.}(2007)\citenamefont {Brion},
  \citenamefont {M\o{}lmer},\ and\ \citenamefont {Saffman}}]{Brion2007quantum}%
  \BibitemOpen
  \bibfield  {author} {\bibinfo {author} {\bibfnamefont {E.}~\bibnamefont
  {Brion}}, \bibinfo {author} {\bibfnamefont {K.}~\bibnamefont {M\o{}lmer}},\
  and\ \bibinfo {author} {\bibfnamefont {M.}~\bibnamefont {Saffman}},\
  }\bibfield  {title} {\bibinfo {title} {Quantum computing with collective
  ensembles of multilevel systems},\ }\href
  {https://doi.org/10.1103/PhysRevLett.99.260501} {\bibfield  {journal}
  {\bibinfo  {journal} {Phys. Rev. Lett.}\ }\textbf {\bibinfo {volume} {99}},\
  \bibinfo {pages} {260501} (\bibinfo {year} {2007})}\BibitemShut {NoStop}%
\bibitem [{\citenamefont {Barrett}\ \emph {et~al.}(2010)\citenamefont
  {Barrett}, \citenamefont {Rohde},\ and\ \citenamefont
  {Stace}}]{Barrett2010scalable}%
  \BibitemOpen
  \bibfield  {author} {\bibinfo {author} {\bibfnamefont {S.~D.}\ \bibnamefont
  {Barrett}}, \bibinfo {author} {\bibfnamefont {P.~P.}\ \bibnamefont {Rohde}},\
  and\ \bibinfo {author} {\bibfnamefont {T.~M.}\ \bibnamefont {Stace}},\
  }\bibfield  {title} {\bibinfo {title} {Scalable quantum computing with atomic
  ensembles},\ }\href {https://doi.org/10.1088/1367-2630/12/9/093032}
  {\bibfield  {journal} {\bibinfo  {journal} {New J. Phys.}\ }\textbf {\bibinfo
  {volume} {12}},\ \bibinfo {pages} {093032} (\bibinfo {year}
  {2010})}\BibitemShut {NoStop}%
\bibitem [{\citenamefont {Georgescu}\ \emph {et~al.}(2014)\citenamefont
  {Georgescu}, \citenamefont {Ashhab},\ and\ \citenamefont
  {Nori}}]{Georgescu2014quantum}%
  \BibitemOpen
  \bibfield  {author} {\bibinfo {author} {\bibfnamefont {I.~M.}\ \bibnamefont
  {Georgescu}}, \bibinfo {author} {\bibfnamefont {S.}~\bibnamefont {Ashhab}},\
  and\ \bibinfo {author} {\bibfnamefont {F.}~\bibnamefont {Nori}},\ }\bibfield
  {title} {\bibinfo {title} {Quantum simulation},\ }\href
  {https://doi.org/10.1103/RevModPhys.86.153} {\bibfield  {journal} {\bibinfo
  {journal} {Rev. Mod. Phys.}\ }\textbf {\bibinfo {volume} {86}},\ \bibinfo
  {pages} {153} (\bibinfo {year} {2014})}\BibitemShut {NoStop}%
\bibitem [{\citenamefont {Islam}\ \emph {et~al.}(2011)\citenamefont {Islam},
  \citenamefont {Edwards}, \citenamefont {Kim}, \citenamefont {Korenblit},
  \citenamefont {Noh}, \citenamefont {Carmichael}, \citenamefont {Lin},
  \citenamefont {Duan}, \citenamefont {Joseph~Wang}, \citenamefont
  {Freericks},\ and\ \citenamefont {Monroe}}]{Islam2011onset}%
  \BibitemOpen
  \bibfield  {author} {\bibinfo {author} {\bibfnamefont {R.}~\bibnamefont
  {Islam}}, \bibinfo {author} {\bibfnamefont {E.~E.}\ \bibnamefont {Edwards}},
  \bibinfo {author} {\bibfnamefont {K.}~\bibnamefont {Kim}}, \bibinfo {author}
  {\bibfnamefont {S.}~\bibnamefont {Korenblit}}, \bibinfo {author}
  {\bibfnamefont {C.}~\bibnamefont {Noh}}, \bibinfo {author} {\bibfnamefont
  {H.}~\bibnamefont {Carmichael}}, \bibinfo {author} {\bibfnamefont {G.-D.}\
  \bibnamefont {Lin}}, \bibinfo {author} {\bibfnamefont {L.-M.}\ \bibnamefont
  {Duan}}, \bibinfo {author} {\bibfnamefont {C.-C.}\ \bibnamefont
  {Joseph~Wang}}, \bibinfo {author} {\bibfnamefont {J.~K.}\ \bibnamefont
  {Freericks}},\ and\ \bibinfo {author} {\bibfnamefont {C.}~\bibnamefont
  {Monroe}},\ }\bibfield  {title} {\bibinfo {title} {Onset of a quantum phase
  transition with a trapped ion quantum simulator},\ }\href
  {https://doi.org/10.1038/ncomms1374} {\bibfield  {journal} {\bibinfo
  {journal} {Nat. Commun.}\ }\textbf {\bibinfo {volume} {2}},\ \bibinfo {pages}
  {377} (\bibinfo {year} {2011})}\BibitemShut {NoStop}%
\bibitem [{\citenamefont {Koppenh\"ofer}\ \emph {et~al.}(2022)\citenamefont
  {Koppenh\"ofer}, \citenamefont {Groszkowski}, \citenamefont {Lau},\ and\
  \citenamefont {Clerk}}]{Koppenhofer2022dissipative}%
  \BibitemOpen
  \bibfield  {author} {\bibinfo {author} {\bibfnamefont {M.}~\bibnamefont
  {Koppenh\"ofer}}, \bibinfo {author} {\bibfnamefont {P.}~\bibnamefont
  {Groszkowski}}, \bibinfo {author} {\bibfnamefont {H.-K.}\ \bibnamefont
  {Lau}},\ and\ \bibinfo {author} {\bibfnamefont {A.}~\bibnamefont {Clerk}},\
  }\bibfield  {title} {\bibinfo {title} {Dissipative superradiant spin
  amplifier for enhanced quantum sensing},\ }\href
  {https://doi.org/10.1103/PRXQuantum.3.030330} {\bibfield  {journal} {\bibinfo
   {journal} {PRX Quantum}\ }\textbf {\bibinfo {volume} {3}},\ \bibinfo {pages}
  {030330} (\bibinfo {year} {2022})}\BibitemShut {NoStop}%
\bibitem [{\citenamefont {Duan}\ \emph {et~al.}(2001)\citenamefont {Duan},
  \citenamefont {Lukin}, \citenamefont {Cirac},\ and\ \citenamefont
  {Zoller}}]{Duan2001long}%
  \BibitemOpen
  \bibfield  {author} {\bibinfo {author} {\bibfnamefont {L.-M.}\ \bibnamefont
  {Duan}}, \bibinfo {author} {\bibfnamefont {M.~D.}\ \bibnamefont {Lukin}},
  \bibinfo {author} {\bibfnamefont {J.~I.}\ \bibnamefont {Cirac}},\ and\
  \bibinfo {author} {\bibfnamefont {P.}~\bibnamefont {Zoller}},\ }\bibfield
  {title} {\bibinfo {title} {Long-distance quantum communication with atomic
  ensembles and linear optics},\ }\href {https://doi.org/10.1038/35106500}
  {\bibfield  {journal} {\bibinfo  {journal} {Nature}\ }\textbf {\bibinfo
  {volume} {414}},\ \bibinfo {pages} {413} (\bibinfo {year}
  {2001})}\BibitemShut {NoStop}%
\bibitem [{\citenamefont {Sangouard}\ \emph {et~al.}(2011)\citenamefont
  {Sangouard}, \citenamefont {Simon}, \citenamefont {de~Riedmatten},\ and\
  \citenamefont {Gisin}}]{Sangouard2011quantum}%
  \BibitemOpen
  \bibfield  {author} {\bibinfo {author} {\bibfnamefont {N.}~\bibnamefont
  {Sangouard}}, \bibinfo {author} {\bibfnamefont {C.}~\bibnamefont {Simon}},
  \bibinfo {author} {\bibfnamefont {H.}~\bibnamefont {de~Riedmatten}},\ and\
  \bibinfo {author} {\bibfnamefont {N.}~\bibnamefont {Gisin}},\ }\bibfield
  {title} {\bibinfo {title} {Quantum repeaters based on atomic ensembles and
  linear optics},\ }\href {https://doi.org/10.1103/RevModPhys.83.33} {\bibfield
   {journal} {\bibinfo  {journal} {Rev. Mod. Phys.}\ }\textbf {\bibinfo
  {volume} {83}},\ \bibinfo {pages} {33} (\bibinfo {year} {2011})}\BibitemShut
  {NoStop}%
\bibitem [{\citenamefont {Dicke}(1954)}]{Dicke1954coherence}%
  \BibitemOpen
  \bibfield  {author} {\bibinfo {author} {\bibfnamefont {R.~H.}\ \bibnamefont
  {Dicke}},\ }\bibfield  {title} {\bibinfo {title} {Coherence in spontaneous
  radiation processes},\ }\href {https://doi.org/10.1103/PhysRev.93.99}
  {\bibfield  {journal} {\bibinfo  {journal} {Phys. Rev.}\ }\textbf {\bibinfo
  {volume} {93}},\ \bibinfo {pages} {99} (\bibinfo {year} {1954})}\BibitemShut
  {NoStop}%
\bibitem [{\citenamefont {Gross}\ and\ \citenamefont
  {Haroche}(1982)}]{Gross1982superradiance}%
  \BibitemOpen
  \bibfield  {author} {\bibinfo {author} {\bibfnamefont {M.}~\bibnamefont
  {Gross}}\ and\ \bibinfo {author} {\bibfnamefont {S.}~\bibnamefont
  {Haroche}},\ }\bibfield  {title} {\bibinfo {title} {Superradiance: An essay
  on the theory of collective spontaneous emission},\ }\href
  {https://doi.org/https://doi.org/10.1016/0370-1573(82)90102-8} {\bibfield
  {journal} {\bibinfo  {journal} {Phys. Rep.}\ }\textbf {\bibinfo {volume}
  {93}},\ \bibinfo {pages} {301} (\bibinfo {year} {1982})}\BibitemShut
  {NoStop}%
\bibitem [{\citenamefont {Xu}\ \emph {et~al.}(2014)\citenamefont {Xu},
  \citenamefont {Tieri}, \citenamefont {Fine}, \citenamefont {Thompson},\ and\
  \citenamefont {Holland}}]{Xu2014synchronization}%
  \BibitemOpen
  \bibfield  {author} {\bibinfo {author} {\bibfnamefont {M.}~\bibnamefont
  {Xu}}, \bibinfo {author} {\bibfnamefont {D.~A.}\ \bibnamefont {Tieri}},
  \bibinfo {author} {\bibfnamefont {E.~C.}\ \bibnamefont {Fine}}, \bibinfo
  {author} {\bibfnamefont {J.~K.}\ \bibnamefont {Thompson}},\ and\ \bibinfo
  {author} {\bibfnamefont {M.~J.}\ \bibnamefont {Holland}},\ }\bibfield
  {title} {\bibinfo {title} {Synchronization of two ensembles of atoms},\
  }\href {https://doi.org/10.1103/PhysRevLett.113.154101} {\bibfield  {journal}
  {\bibinfo  {journal} {Phys. Rev. Lett.}\ }\textbf {\bibinfo {volume} {113}},\
  \bibinfo {pages} {154101} (\bibinfo {year} {2014})}\BibitemShut {NoStop}%
\bibitem [{\citenamefont {Masson}\ \emph {et~al.}(2020)\citenamefont {Masson},
  \citenamefont {Ferrier-Barbut}, \citenamefont {Orozco}, \citenamefont
  {Browaeys},\ and\ \citenamefont {Asenjo-Garcia}}]{Masson2020many}%
  \BibitemOpen
  \bibfield  {author} {\bibinfo {author} {\bibfnamefont {S.~J.}\ \bibnamefont
  {Masson}}, \bibinfo {author} {\bibfnamefont {I.}~\bibnamefont
  {Ferrier-Barbut}}, \bibinfo {author} {\bibfnamefont {L.~A.}\ \bibnamefont
  {Orozco}}, \bibinfo {author} {\bibfnamefont {A.}~\bibnamefont {Browaeys}},\
  and\ \bibinfo {author} {\bibfnamefont {A.}~\bibnamefont {Asenjo-Garcia}},\
  }\bibfield  {title} {\bibinfo {title} {Many-body signatures of collective
  decay in atomic chains},\ }\href
  {https://doi.org/10.1103/PhysRevLett.125.263601} {\bibfield  {journal}
  {\bibinfo  {journal} {Phys. Rev. Lett.}\ }\textbf {\bibinfo {volume} {125}},\
  \bibinfo {pages} {263601} (\bibinfo {year} {2020})}\BibitemShut {NoStop}%
\bibitem [{\citenamefont {Masson}\ and\ \citenamefont
  {Asenjo-Garcia}(2022)}]{Masson2022universality}%
  \BibitemOpen
  \bibfield  {author} {\bibinfo {author} {\bibfnamefont {S.~J.}\ \bibnamefont
  {Masson}}\ and\ \bibinfo {author} {\bibfnamefont {A.}~\bibnamefont
  {Asenjo-Garcia}},\ }\bibfield  {title} {\bibinfo {title} {Universality of
  dicke superradiance in arrays of quantum emitters},\ }\href
  {https://doi.org/10.1038/s41467-022-29805-4} {\bibfield  {journal} {\bibinfo
  {journal} {Nat. Commun.}\ }\textbf {\bibinfo {volume} {13}},\ \bibinfo
  {pages} {2285} (\bibinfo {year} {2022})}\BibitemShut {NoStop}%
\bibitem [{\citenamefont {Sierra}\ \emph {et~al.}(2022)\citenamefont {Sierra},
  \citenamefont {Masson},\ and\ \citenamefont
  {Asenjo-Garcia}}]{Sierra2021dicke}%
  \BibitemOpen
  \bibfield  {author} {\bibinfo {author} {\bibfnamefont {E.}~\bibnamefont
  {Sierra}}, \bibinfo {author} {\bibfnamefont {S.~J.}\ \bibnamefont {Masson}},\
  and\ \bibinfo {author} {\bibfnamefont {A.}~\bibnamefont {Asenjo-Garcia}},\
  }\bibfield  {title} {\bibinfo {title} {Dicke superradiance in ordered
  lattices: Dimensionality matters},\ }\href
  {https://doi.org/10.1103/PhysRevResearch.4.023207} {\bibfield  {journal}
  {\bibinfo  {journal} {Phys. Rev. Research}\ }\textbf {\bibinfo {volume}
  {4}},\ \bibinfo {pages} {023207} (\bibinfo {year} {2022})}\BibitemShut
  {NoStop}%
\bibitem [{\citenamefont {Robicheaux}(2021)}]{Robicheaux2021theoretical}%
  \BibitemOpen
  \bibfield  {author} {\bibinfo {author} {\bibfnamefont {F.}~\bibnamefont
  {Robicheaux}},\ }\bibfield  {title} {\bibinfo {title} {Theoretical study of
  early-time superradiance for atom clouds and arrays},\ }\href
  {https://doi.org/10.1103/PhysRevA.104.063706} {\bibfield  {journal} {\bibinfo
   {journal} {Phys. Rev. A}\ }\textbf {\bibinfo {volume} {104}},\ \bibinfo
  {pages} {063706} (\bibinfo {year} {2021})}\BibitemShut {NoStop}%
\bibitem [{\citenamefont {Malz}\ \emph {et~al.}(2022)\citenamefont {Malz},
  \citenamefont {Trivedi},\ and\ \citenamefont {Cirac}}]{Malz2022large}%
  \BibitemOpen
  \bibfield  {author} {\bibinfo {author} {\bibfnamefont {D.}~\bibnamefont
  {Malz}}, \bibinfo {author} {\bibfnamefont {R.}~\bibnamefont {Trivedi}},\ and\
  \bibinfo {author} {\bibfnamefont {J.~I.}\ \bibnamefont {Cirac}},\ }\bibfield
  {title} {\bibinfo {title} {Large-$n$ limit of dicke superradiance},\ }\href
  {https://doi.org/10.1103/PhysRevA.106.013716} {\bibfield  {journal} {\bibinfo
   {journal} {Phys. Rev. A}\ }\textbf {\bibinfo {volume} {106}},\ \bibinfo
  {pages} {013716} (\bibinfo {year} {2022})}\BibitemShut {NoStop}%
\bibitem [{\citenamefont {Cardenas-Lopez}\ \emph {et~al.}(2023)\citenamefont
  {Cardenas-Lopez}, \citenamefont {Masson}, \citenamefont {Zager},\ and\
  \citenamefont {Asenjo-Garcia}}]{Silvia2023many}%
  \BibitemOpen
  \bibfield  {author} {\bibinfo {author} {\bibfnamefont {S.}~\bibnamefont
  {Cardenas-Lopez}}, \bibinfo {author} {\bibfnamefont {S.~J.}\ \bibnamefont
  {Masson}}, \bibinfo {author} {\bibfnamefont {Z.}~\bibnamefont {Zager}},\ and\
  \bibinfo {author} {\bibfnamefont {A.}~\bibnamefont {Asenjo-Garcia}},\
  }\bibfield  {title} {\bibinfo {title} {Many-body superradiance and dynamical
  mirror symmetry breaking in waveguide qed},\ }\href
  {https://doi.org/10.1103/PhysRevLett.131.033605} {\bibfield  {journal}
  {\bibinfo  {journal} {Phys. Rev. Lett.}\ }\textbf {\bibinfo {volume} {131}},\
  \bibinfo {pages} {033605} (\bibinfo {year} {2023})}\BibitemShut {NoStop}%
\bibitem [{\citenamefont {Mok}\ \emph {et~al.}(2023)\citenamefont {Mok},
  \citenamefont {Asenjo-Garcia}, \citenamefont {Sum},\ and\ \citenamefont
  {Kwek}}]{Mok2023dicke}%
  \BibitemOpen
  \bibfield  {author} {\bibinfo {author} {\bibfnamefont {W.-K.}\ \bibnamefont
  {Mok}}, \bibinfo {author} {\bibfnamefont {A.}~\bibnamefont {Asenjo-Garcia}},
  \bibinfo {author} {\bibfnamefont {T.~C.}\ \bibnamefont {Sum}},\ and\ \bibinfo
  {author} {\bibfnamefont {L.-C.}\ \bibnamefont {Kwek}},\ }\bibfield  {title}
  {\bibinfo {title} {Dicke superradiance requires interactions beyond nearest
  neighbors},\ }\href {https://doi.org/10.1103/PhysRevLett.130.213605}
  {\bibfield  {journal} {\bibinfo  {journal} {Phys. Rev. Lett.}\ }\textbf
  {\bibinfo {volume} {130}},\ \bibinfo {pages} {213605} (\bibinfo {year}
  {2023})}\BibitemShut {NoStop}%
\bibitem [{\citenamefont {Rainò}\ \emph {et~al.}(2020)\citenamefont {Rainò},
  \citenamefont {Utzat}, \citenamefont {Bawendi},\ and\ \citenamefont
  {Kovalenko}}]{Raino2020superradiant}%
  \BibitemOpen
  \bibfield  {author} {\bibinfo {author} {\bibfnamefont {G.}~\bibnamefont
  {Rainò}}, \bibinfo {author} {\bibfnamefont {H.}~\bibnamefont {Utzat}},
  \bibinfo {author} {\bibfnamefont {M.}~\bibnamefont {Bawendi}},\ and\ \bibinfo
  {author} {\bibfnamefont {M.}~\bibnamefont {Kovalenko}},\ }\bibfield  {title}
  {\bibinfo {title} {Superradiant emission from self-assembled light emitters:
  From molecules to quantum dots},\ }\href
  {https://doi.org/10.1557/mrs.2020.250} {\bibfield  {journal} {\bibinfo
  {journal} {MRS Bulletin}\ }\textbf {\bibinfo {volume} {45}},\ \bibinfo
  {pages} {841–848} (\bibinfo {year} {2020})}\BibitemShut {NoStop}%
\bibitem [{\citenamefont {Rain{\`o}}\ \emph {et~al.}(2018)\citenamefont
  {Rain{\`o}}, \citenamefont {Becker}, \citenamefont {Bodnarchuk},
  \citenamefont {Mahrt}, \citenamefont {Kovalenko},\ and\ \citenamefont
  {St{\"o}ferle}}]{Raino2018superfluorescence}%
  \BibitemOpen
  \bibfield  {author} {\bibinfo {author} {\bibfnamefont {G.}~\bibnamefont
  {Rain{\`o}}}, \bibinfo {author} {\bibfnamefont {M.~A.}\ \bibnamefont
  {Becker}}, \bibinfo {author} {\bibfnamefont {M.~I.}\ \bibnamefont
  {Bodnarchuk}}, \bibinfo {author} {\bibfnamefont {R.~F.}\ \bibnamefont
  {Mahrt}}, \bibinfo {author} {\bibfnamefont {M.~V.}\ \bibnamefont
  {Kovalenko}},\ and\ \bibinfo {author} {\bibfnamefont {T.}~\bibnamefont
  {St{\"o}ferle}},\ }\bibfield  {title} {\bibinfo {title} {Superfluorescence
  from lead halide perovskite quantum dot superlattices},\ }\href
  {https://doi.org/10.1038/s41586-018-0683-0} {\bibfield  {journal} {\bibinfo
  {journal} {Nature}\ }\textbf {\bibinfo {volume} {563}},\ \bibinfo {pages}
  {671} (\bibinfo {year} {2018})}\BibitemShut {NoStop}%
\bibitem [{\citenamefont {Lei}\ \emph {et~al.}(2023)\citenamefont {Lei},
  \citenamefont {Fukumori}, \citenamefont {Rochman}, \citenamefont {Zhu},
  \citenamefont {Endres}, \citenamefont {Choi},\ and\ \citenamefont
  {Faraon}}]{Lei2023many}%
  \BibitemOpen
  \bibfield  {author} {\bibinfo {author} {\bibfnamefont {M.}~\bibnamefont
  {Lei}}, \bibinfo {author} {\bibfnamefont {R.}~\bibnamefont {Fukumori}},
  \bibinfo {author} {\bibfnamefont {J.}~\bibnamefont {Rochman}}, \bibinfo
  {author} {\bibfnamefont {B.}~\bibnamefont {Zhu}}, \bibinfo {author}
  {\bibfnamefont {M.}~\bibnamefont {Endres}}, \bibinfo {author} {\bibfnamefont
  {J.}~\bibnamefont {Choi}},\ and\ \bibinfo {author} {\bibfnamefont
  {A.}~\bibnamefont {Faraon}},\ }\bibfield  {title} {\bibinfo {title}
  {Many-body cavity quantum electrodynamics with driven inhomogeneous
  emitters},\ }\href {https://doi.org/10.1038/s41586-023-05884-1} {\bibfield
  {journal} {\bibinfo  {journal} {Nature}\ }\textbf {\bibinfo {volume} {617}},\
  \bibinfo {pages} {271} (\bibinfo {year} {2023})}\BibitemShut {NoStop}%
\bibitem [{\citenamefont {Leghtas}\ \emph {et~al.}(2015)\citenamefont
  {Leghtas}, \citenamefont {Touzard}, \citenamefont {Pop}, \citenamefont {Kou},
  \citenamefont {Vlastakis}, \citenamefont {Petrenko}, \citenamefont {Sliwa},
  \citenamefont {Narla}, \citenamefont {Shankar}, \citenamefont {Hatridge},
  \citenamefont {Reagor}, \citenamefont {Frunzio}, \citenamefont {Schoelkopf},
  \citenamefont {Mirrahimi},\ and\ \citenamefont
  {Devoret}}]{Leghtas2015confining}%
  \BibitemOpen
  \bibfield  {author} {\bibinfo {author} {\bibfnamefont {Z.}~\bibnamefont
  {Leghtas}}, \bibinfo {author} {\bibfnamefont {S.}~\bibnamefont {Touzard}},
  \bibinfo {author} {\bibfnamefont {I.~M.}\ \bibnamefont {Pop}}, \bibinfo
  {author} {\bibfnamefont {A.}~\bibnamefont {Kou}}, \bibinfo {author}
  {\bibfnamefont {B.}~\bibnamefont {Vlastakis}}, \bibinfo {author}
  {\bibfnamefont {A.}~\bibnamefont {Petrenko}}, \bibinfo {author}
  {\bibfnamefont {K.~M.}\ \bibnamefont {Sliwa}}, \bibinfo {author}
  {\bibfnamefont {A.}~\bibnamefont {Narla}}, \bibinfo {author} {\bibfnamefont
  {S.}~\bibnamefont {Shankar}}, \bibinfo {author} {\bibfnamefont {M.~J.}\
  \bibnamefont {Hatridge}}, \bibinfo {author} {\bibfnamefont {M.}~\bibnamefont
  {Reagor}}, \bibinfo {author} {\bibfnamefont {L.}~\bibnamefont {Frunzio}},
  \bibinfo {author} {\bibfnamefont {R.~J.}\ \bibnamefont {Schoelkopf}},
  \bibinfo {author} {\bibfnamefont {M.}~\bibnamefont {Mirrahimi}},\ and\
  \bibinfo {author} {\bibfnamefont {M.~H.}\ \bibnamefont {Devoret}},\
  }\bibfield  {title} {\bibinfo {title} {Confining the state of light to a
  quantum manifold by engineered two-photon loss},\ }\href
  {https://doi.org/10.1126/science.aaa2085} {\bibfield  {journal} {\bibinfo
  {journal} {Science}\ }\textbf {\bibinfo {volume} {347}},\ \bibinfo {pages}
  {853} (\bibinfo {year} {2015})}\BibitemShut {NoStop}%
\bibitem [{\citenamefont {Lescanne}\ \emph {et~al.}(2020)\citenamefont
  {Lescanne}, \citenamefont {Villiers}, \citenamefont {Peronnin}, \citenamefont
  {Sarlette}, \citenamefont {Delbecq}, \citenamefont {Huard}, \citenamefont
  {Kontos}, \citenamefont {Mirrahimi},\ and\ \citenamefont
  {Leghtas}}]{Lescanne2020exponential}%
  \BibitemOpen
  \bibfield  {author} {\bibinfo {author} {\bibfnamefont {R.}~\bibnamefont
  {Lescanne}}, \bibinfo {author} {\bibfnamefont {M.}~\bibnamefont {Villiers}},
  \bibinfo {author} {\bibfnamefont {T.}~\bibnamefont {Peronnin}}, \bibinfo
  {author} {\bibfnamefont {A.}~\bibnamefont {Sarlette}}, \bibinfo {author}
  {\bibfnamefont {M.}~\bibnamefont {Delbecq}}, \bibinfo {author} {\bibfnamefont
  {B.}~\bibnamefont {Huard}}, \bibinfo {author} {\bibfnamefont
  {T.}~\bibnamefont {Kontos}}, \bibinfo {author} {\bibfnamefont
  {M.}~\bibnamefont {Mirrahimi}},\ and\ \bibinfo {author} {\bibfnamefont
  {Z.}~\bibnamefont {Leghtas}},\ }\bibfield  {title} {\bibinfo {title}
  {Exponential suppression of bit-flips in a qubit encoded in an oscillator},\
  }\href {https://doi.org/10.1038/s41567-020-0824-x} {\bibfield  {journal}
  {\bibinfo  {journal} {Nat. Phys.}\ }\textbf {\bibinfo {volume} {16}},\
  \bibinfo {pages} {509} (\bibinfo {year} {2020})}\BibitemShut {NoStop}%
\bibitem [{\citenamefont {Radcliffe}(1971)}]{Radcliffe1971some}%
  \BibitemOpen
  \bibfield  {author} {\bibinfo {author} {\bibfnamefont {J.~M.}\ \bibnamefont
  {Radcliffe}},\ }\bibfield  {title} {\bibinfo {title} {Some properties of
  coherent spin states},\ }\href {https://doi.org/10.1088/0305-4470/4/3/009}
  {\bibfield  {journal} {\bibinfo  {journal} {J. Phys. A: Gen. Phys.}\ }\textbf
  {\bibinfo {volume} {4}},\ \bibinfo {pages} {313} (\bibinfo {year}
  {1971})}\BibitemShut {NoStop}%
\bibitem [{\citenamefont {Qin}\ \emph {et~al.}(2021)\citenamefont {Qin},
  \citenamefont {Miranowicz}, \citenamefont {Jing},\ and\ \citenamefont
  {Nori}}]{qin2021generating}%
  \BibitemOpen
  \bibfield  {author} {\bibinfo {author} {\bibfnamefont {W.}~\bibnamefont
  {Qin}}, \bibinfo {author} {\bibfnamefont {A.}~\bibnamefont {Miranowicz}},
  \bibinfo {author} {\bibfnamefont {H.}~\bibnamefont {Jing}},\ and\ \bibinfo
  {author} {\bibfnamefont {F.}~\bibnamefont {Nori}},\ }\bibfield  {title}
  {\bibinfo {title} {Generating long-lived macroscopically distinct
  superposition states in atomic ensembles},\ }\href
  {https://doi.org/10.1103/PhysRevLett.127.093602} {\bibfield  {journal}
  {\bibinfo  {journal} {Phys. Rev. Lett.}\ }\textbf {\bibinfo {volume} {127}},\
  \bibinfo {pages} {093602} (\bibinfo {year} {2021})}\BibitemShut {NoStop}%
\bibitem [{\citenamefont {Mirrahimi}\ \emph {et~al.}(2014)\citenamefont
  {Mirrahimi}, \citenamefont {Leghtas}, \citenamefont {Albert}, \citenamefont
  {Touzard}, \citenamefont {Schoelkopf}, \citenamefont {Jiang},\ and\
  \citenamefont {Devoret}}]{Mirrahimi2014dynamically}%
  \BibitemOpen
  \bibfield  {author} {\bibinfo {author} {\bibfnamefont {M.}~\bibnamefont
  {Mirrahimi}}, \bibinfo {author} {\bibfnamefont {Z.}~\bibnamefont {Leghtas}},
  \bibinfo {author} {\bibfnamefont {V.~V.}\ \bibnamefont {Albert}}, \bibinfo
  {author} {\bibfnamefont {S.}~\bibnamefont {Touzard}}, \bibinfo {author}
  {\bibfnamefont {R.~J.}\ \bibnamefont {Schoelkopf}}, \bibinfo {author}
  {\bibfnamefont {L.}~\bibnamefont {Jiang}},\ and\ \bibinfo {author}
  {\bibfnamefont {M.~H.}\ \bibnamefont {Devoret}},\ }\bibfield  {title}
  {\bibinfo {title} {Dynamically protected cat-qubits: a new paradigm for
  universal quantum computation},\ }\href
  {https://doi.org/10.1088/1367-2630/16/4/045014} {\bibfield  {journal}
  {\bibinfo  {journal} {New J. Phys.}\ }\textbf {\bibinfo {volume} {16}},\
  \bibinfo {pages} {045014} (\bibinfo {year} {2014})}\BibitemShut {NoStop}%
\bibitem [{\citenamefont {Johansson}\ \emph {et~al.}(2013)\citenamefont
  {Johansson}, \citenamefont {Nation},\ and\ \citenamefont
  {Nori}}]{Johansson2013qutip}%
  \BibitemOpen
  \bibfield  {author} {\bibinfo {author} {\bibfnamefont {J.}~\bibnamefont
  {Johansson}}, \bibinfo {author} {\bibfnamefont {P.}~\bibnamefont {Nation}},\
  and\ \bibinfo {author} {\bibfnamefont {F.}~\bibnamefont {Nori}},\ }\bibfield
  {title} {\bibinfo {title} {{QuTiP} 2: A python framework for the dynamics of
  open quantum systems},\ }\href {https://doi.org/10.1016/j.cpc.2012.11.019}
  {\bibfield  {journal} {\bibinfo  {journal} {Comput. Phys. Commun.}\ }\textbf
  {\bibinfo {volume} {184}},\ \bibinfo {pages} {1234} (\bibinfo {year}
  {2013})}\BibitemShut {NoStop}%
\bibitem [{\citenamefont {Leghtas}\ \emph {et~al.}(2013)\citenamefont
  {Leghtas}, \citenamefont {Kirchmair}, \citenamefont {Vlastakis},
  \citenamefont {Schoelkopf}, \citenamefont {Devoret},\ and\ \citenamefont
  {Mirrahimi}}]{Leghtas2013hardware}%
  \BibitemOpen
  \bibfield  {author} {\bibinfo {author} {\bibfnamefont {Z.}~\bibnamefont
  {Leghtas}}, \bibinfo {author} {\bibfnamefont {G.}~\bibnamefont {Kirchmair}},
  \bibinfo {author} {\bibfnamefont {B.}~\bibnamefont {Vlastakis}}, \bibinfo
  {author} {\bibfnamefont {R.~J.}\ \bibnamefont {Schoelkopf}}, \bibinfo
  {author} {\bibfnamefont {M.~H.}\ \bibnamefont {Devoret}},\ and\ \bibinfo
  {author} {\bibfnamefont {M.}~\bibnamefont {Mirrahimi}},\ }\bibfield  {title}
  {\bibinfo {title} {Hardware-efficient autonomous quantum memory protection},\
  }\href {https://doi.org/10.1103/PhysRevLett.111.120501} {\bibfield  {journal}
  {\bibinfo  {journal} {Phys. Rev. Lett.}\ }\textbf {\bibinfo {volume} {111}},\
  \bibinfo {pages} {120501} (\bibinfo {year} {2013})}\BibitemShut {NoStop}%
\bibitem [{\citenamefont {Kubo}(1962)}]{kubo1962generalized}%
  \BibitemOpen
  \bibfield  {author} {\bibinfo {author} {\bibfnamefont {R.}~\bibnamefont
  {Kubo}},\ }\bibfield  {title} {\bibinfo {title} {Generalized cumulant
  expansion method},\ }\href {https://doi.org/10.1143/JPSJ.17.1100} {\bibfield
  {journal} {\bibinfo  {journal} {J. Phys. Soc. Jpn.}\ }\textbf {\bibinfo
  {volume} {17}},\ \bibinfo {pages} {1100} (\bibinfo {year}
  {1962})}\BibitemShut {NoStop}%
\bibitem [{\citenamefont {Plankensteiner}\ \emph {et~al.}(2022)\citenamefont
  {Plankensteiner}, \citenamefont {Hotter},\ and\ \citenamefont
  {Ritsch}}]{Plankensteiner2022quantumcumulantsjl}%
  \BibitemOpen
  \bibfield  {author} {\bibinfo {author} {\bibfnamefont {D.}~\bibnamefont
  {Plankensteiner}}, \bibinfo {author} {\bibfnamefont {C.}~\bibnamefont
  {Hotter}},\ and\ \bibinfo {author} {\bibfnamefont {H.}~\bibnamefont
  {Ritsch}},\ }\bibfield  {title} {\bibinfo {title} {Quantum{C}umulants.jl: {A}
  {J}ulia framework for generalized mean-field equations in open quantum
  systems},\ }\href {https://doi.org/10.22331/q-2022-01-04-617} {\bibfield
  {journal} {\bibinfo  {journal} {{Quantum}}\ }\textbf {\bibinfo {volume}
  {6}},\ \bibinfo {pages} {617} (\bibinfo {year} {2022})}\BibitemShut {NoStop}%
\bibitem [{\citenamefont {Hotter}\ \emph {et~al.}(2023)\citenamefont {Hotter},
  \citenamefont {Ostermann},\ and\ \citenamefont {Ritsch}}]{Hotter2023cavity}%
  \BibitemOpen
  \bibfield  {author} {\bibinfo {author} {\bibfnamefont {C.}~\bibnamefont
  {Hotter}}, \bibinfo {author} {\bibfnamefont {L.}~\bibnamefont {Ostermann}},\
  and\ \bibinfo {author} {\bibfnamefont {H.}~\bibnamefont {Ritsch}},\
  }\bibfield  {title} {\bibinfo {title} {Cavity sub- and superradiance for
  transversely driven atomic ensembles},\ }\href
  {https://doi.org/10.1103/PhysRevResearch.5.013056} {\bibfield  {journal}
  {\bibinfo  {journal} {Phys. Rev. Res.}\ }\textbf {\bibinfo {volume} {5}},\
  \bibinfo {pages} {013056} (\bibinfo {year} {2023})}\BibitemShut {NoStop}%
\bibitem [{\citenamefont {Debnath}\ \emph {et~al.}(2018)\citenamefont
  {Debnath}, \citenamefont {Zhang},\ and\ \citenamefont
  {M\o{}lmer}}]{debnath2018lasing}%
  \BibitemOpen
  \bibfield  {author} {\bibinfo {author} {\bibfnamefont {K.}~\bibnamefont
  {Debnath}}, \bibinfo {author} {\bibfnamefont {Y.}~\bibnamefont {Zhang}},\
  and\ \bibinfo {author} {\bibfnamefont {K.}~\bibnamefont {M\o{}lmer}},\
  }\bibfield  {title} {\bibinfo {title} {Lasing in the superradiant crossover
  regime},\ }\href {https://doi.org/10.1103/PhysRevA.98.063837} {\bibfield
  {journal} {\bibinfo  {journal} {Phys. Rev. A}\ }\textbf {\bibinfo {volume}
  {98}},\ \bibinfo {pages} {063837} (\bibinfo {year} {2018})}\BibitemShut
  {NoStop}%
\bibitem [{\citenamefont {Debnath}\ \emph {et~al.}(2019)\citenamefont
  {Debnath}, \citenamefont {Zhang},\ and\ \citenamefont
  {M\o{}lmer}}]{debnath2019collective}%
  \BibitemOpen
  \bibfield  {author} {\bibinfo {author} {\bibfnamefont {K.}~\bibnamefont
  {Debnath}}, \bibinfo {author} {\bibfnamefont {Y.}~\bibnamefont {Zhang}},\
  and\ \bibinfo {author} {\bibfnamefont {K.}~\bibnamefont {M\o{}lmer}},\
  }\bibfield  {title} {\bibinfo {title} {Collective dynamics of inhomogeneously
  broadened emitters coupled to an optical cavity with narrow linewidth},\
  }\href {https://doi.org/10.1103/PhysRevA.100.053821} {\bibfield  {journal}
  {\bibinfo  {journal} {Phys. Rev. A}\ }\textbf {\bibinfo {volume} {100}},\
  \bibinfo {pages} {053821} (\bibinfo {year} {2019})}\BibitemShut {NoStop}%
\bibitem [{\citenamefont {Debnath}\ \emph {et~al.}(2020)\citenamefont
  {Debnath}, \citenamefont {Dold}, \citenamefont {Morton},\ and\ \citenamefont
  {M\o{}lmer}}]{debnath2020self}%
  \BibitemOpen
  \bibfield  {author} {\bibinfo {author} {\bibfnamefont {K.}~\bibnamefont
  {Debnath}}, \bibinfo {author} {\bibfnamefont {G.}~\bibnamefont {Dold}},
  \bibinfo {author} {\bibfnamefont {J.~J.~L.}\ \bibnamefont {Morton}},\ and\
  \bibinfo {author} {\bibfnamefont {K.}~\bibnamefont {M\o{}lmer}},\ }\bibfield
  {title} {\bibinfo {title} {Self-stimulated pulse echo trains from
  inhomogeneously broadened spin ensembles},\ }\href
  {https://doi.org/10.1103/PhysRevLett.125.137702} {\bibfield  {journal}
  {\bibinfo  {journal} {Phys. Rev. Lett.}\ }\textbf {\bibinfo {volume} {125}},\
  \bibinfo {pages} {137702} (\bibinfo {year} {2020})}\BibitemShut {NoStop}%
\bibitem [{\citenamefont {Rubies-Bigorda}\ \emph {et~al.}(2023)\citenamefont
  {Rubies-Bigorda}, \citenamefont {Ostermann},\ and\ \citenamefont
  {Yelin}}]{rubies2023characterizing}%
  \BibitemOpen
  \bibfield  {author} {\bibinfo {author} {\bibfnamefont {O.}~\bibnamefont
  {Rubies-Bigorda}}, \bibinfo {author} {\bibfnamefont {S.}~\bibnamefont
  {Ostermann}},\ and\ \bibinfo {author} {\bibfnamefont {S.~F.}\ \bibnamefont
  {Yelin}},\ }\bibfield  {title} {\bibinfo {title} {Characterizing superradiant
  dynamics in atomic arrays via a cumulant expansion approach},\ }\href
  {https://doi.org/10.1103/PhysRevResearch.5.013091} {\bibfield  {journal}
  {\bibinfo  {journal} {Phys. Rev. Res.}\ }\textbf {\bibinfo {volume} {5}},\
  \bibinfo {pages} {013091} (\bibinfo {year} {2023})}\BibitemShut {NoStop}%
\bibitem [{\citenamefont {Masson}\ \emph {et~al.}(2024)\citenamefont {Masson},
  \citenamefont {Covey}, \citenamefont {Will},\ and\ \citenamefont
  {Asenjo-Garcia}}]{masson2024dicke}%
  \BibitemOpen
  \bibfield  {author} {\bibinfo {author} {\bibfnamefont {S.~J.}\ \bibnamefont
  {Masson}}, \bibinfo {author} {\bibfnamefont {J.~P.}\ \bibnamefont {Covey}},
  \bibinfo {author} {\bibfnamefont {S.}~\bibnamefont {Will}},\ and\ \bibinfo
  {author} {\bibfnamefont {A.}~\bibnamefont {Asenjo-Garcia}},\ }\bibfield
  {title} {\bibinfo {title} {Dicke superradiance in ordered arrays of
  multilevel atoms},\ }\href {https://doi.org/10.1103/PRXQuantum.5.010344}
  {\bibfield  {journal} {\bibinfo  {journal} {PRX Quantum}\ }\textbf {\bibinfo
  {volume} {5}},\ \bibinfo {pages} {010344} (\bibinfo {year}
  {2024})}\BibitemShut {NoStop}%
\end{thebibliography}%

\newpage
\appendix
\onecolumngrid

\section{Derivation of the spin master equation}
\label{app:derivationME}

Here, we provide a physical derivation of the master equation~\eqref{eq:ME} in the main text. Our approach provides an alternative to that in Ref.~\cite{qin2021generating}, based on time-averaging of the Hamiltonian. 

Consider the Hamiltonian
\begin{equation}
    H = \sum_{n=1}^{N} \frac{\omega_0 + \delta_n}{2} \sigma_n^z + 2 \omega_0 a^\dag a + g \sum_{n=1}^{N} (a^\dag \sigma_n^- + a \sigma_n^+) + \chi^{(2)} (a + a^\dag)^3, 
\end{equation}
which describes the coupling between the spins and the cavity mode with bosonic annihilation operator $a$, in the presence of a nonlinear medium with a large second-order susceptibility $\chi^{(2)}$. The cavity is assumed to be strongly dissipative, i.e., low quality factor. In the dispersive regime where the cavity-spin detuning $\Delta \approx \omega_0$ is large, i.e., $g/\Delta \ll 1$, we have $a \to a + (g/\Delta) \sum_n \sigma_n^- + O(g^2/\Delta^2)$, and the Hamiltonian can be approximated by (in the rotating frame at frequencies $2\omega_0$ with respect to the cavity and $\omega_0$ with respect to the spins)
\begin{equation}
    H \approx \sum_{n=1}^{N} \frac{\delta_n}{2} \sigma_n^z + g_2 \sum_{m,n=1}^N (a \sigma_m^- \sigma_n^- + a^\dag \sigma_m^+ \sigma_n^+)
\end{equation}
where
\begin{equation}
    g_2 = \chi^{(2)} \left(\frac{g}{\Delta}\right)^2.
\end{equation}
Note that $\chi^{(2)}$ must be sufficiently large such that the omitted terms are negligible. Next, we go into the displaced frame $a \to a + \alpha_d$, which can be realized by applying a classical drive with amplitude $\alpha_d$ to the cavity. This gives
\begin{equation}
    H = \sum_{n=1}^{N} \frac{\delta_n}{2} \sigma_n^z + g_2 \sum_{m,n=1}^{N} (\alpha_d \sigma_m^- \sigma_n^- + \alpha_d^* \sigma_m^+ \sigma_n^+) + g_2 \sum_{m,n=1}^{N} (a \sigma_m^- \sigma_n^- + a^\dag \sigma_m^+ \sigma_n^+).
\end{equation}
In the regime where the dissipation rate of the cavity $\kappa$ is much larger than $g_2$, the cavity mode can be adiabatically eliminated. This results in an effective master equation describing the driven-dissipative dynamics of the spins mediated by the cavity:
\begin{equation}
    \dot{\rho} = -i[H_\text{spin},\rho] + \frac{4g_2^2}{\kappa} \mathcal{D}[\sum_{m,n} \sigma_m^- \sigma_n^-]\rho
\end{equation}
with the spin Hamiltonian
\begin{equation}
    H_{\text{spin}} = \frac{1}{2}\sum_{n=1}^{N} \delta_n \sigma_n^z + g_2 \sum_{m,n=1}^{N} (\alpha_d \sigma_m^- \sigma_n^- + \alpha_d^* \sigma_m^+ \sigma_n^+).
\end{equation}
This corresponds to the master equation~\eqref{eq:ME} in the main text, with parameters
\begin{equation}
    \frac{\eta e^{i\varphi}}{N} = g_2 \alpha_d,
\end{equation}
and
\begin{equation}
   \frac{\Gamma_2}{N^2} = \frac{4g_2^2}{\kappa}. 
\end{equation}
The cat amplitude is
\begin{equation}
    |\alpha|^2 = \frac{2\eta}{\Gamma_2} = \frac{\alpha_d \kappa}{2 g_2 N}.
\end{equation}
This implies that the classical drive amplitude $\alpha_d \sim N$ should scale proportionally to the number of spins, in order to keep the cat amplitude fixed.

\section{Derivation of the semiclassical mean-field equations}
\label{app:meanfield}
Starting from the quantum master equation~\eqref{eq:ME}, we make the mean-field ansatz $\rho = \bigotimes_{i=1}^{N} \rho_i$ which assumes that the quantum correlations are negligible. This allows us to derive the single-spin dynamics for $\braket{\sigma_m^+} \equiv \text{Tr}(\sigma_m^+ \rho_m)$ and $\braket{\sigma_m^z} \equiv \text{Tr}(\sigma_m^z \rho_m)$. Equivalently, we factorize spin correlations $\braket{\sigma_i^+ \sigma_j^-} \approx \braket{\sigma_i^+}\braket{\sigma_j^-}$ and $\braket{\sigma_i^+ \sigma_i^- \sigma_j^-} \approx \braket{\sigma_i^+ \sigma_i^-} \braket{\sigma_j^-} = (1 + \braket{\sigma_i^z})\braket{\sigma_j^-}/2$ for $i \neq j$. Note that we treat the population and coherence separately, i.e., we do not assume $\braket{\sigma_i^+ \sigma_i^-} \approx \braket{\sigma_i^+} \braket{\sigma_i^-}$, which allows us to capture some effects of quantum coherence while the spins are weakly excited. The contributions from the squeezing Hamiltonian are
\begin{equation}
    \frac{d}{dt} \braket{\sigma_m^+} = i\frac{\eta}{N} \sum_{i,j} e^{i\theta} \braket{ [\sigma_i^- \sigma_j^- , \sigma_m^+] } + e^{-i\theta} \cancel{ \braket{ [\sigma_i^+ \sigma_j^+ , \sigma_m^+] } } = \frac{2 i \eta}{N} e^{i\theta} \sum_{j\neq m} \braket{ [\sigma_m^-,\sigma_m^+] \sigma_j^- } \approx \frac{-2i\eta}{N} e^{i\theta} \braket{\sigma_m^z} \sum_{j\neq m} \braket{\sigma_j^-}
\label{eq:squeezing_sigma+}
\end{equation}
where the factor of 2 comes from the symmetry of $i \leftrightarrow j$. For $\braket{\sigma_m^z}$, we have
\begin{equation}
\begin{split}
   \frac{d}{dt} \braket{\sigma_m^z} &= \frac{i\eta}{N} \left\{ \sum_{i,j} e^{i\theta} \braket{[\sigma_i^- \sigma_j^-,\sigma_m^z]} + e^{-i\theta} \braket{[\sigma_i^+ \sigma_j^+, \sigma_m^z]}  \right\} = \frac{2i\eta}{N} \left\{ \sum_{j\neq m} e^{i\theta} \braket{[\sigma_m^-,\sigma_m^z]\sigma_j^-} + e^{-i\theta} \braket{[\sigma_m^+, \sigma_m^z]\sigma_j^+}  \right\} \\
   &\approx \frac{4i\eta}{N} \sum_{j\neq m} e^{i\theta} \braket{\sigma_m^-}\braket{\sigma_j^-}-e^{-i\theta}\braket{\sigma_m^+}\braket{\sigma_j^+} = \frac{8\eta}{N} \text{Im} \left( e^{-i\theta} \braket{\sigma_m^+}\sum_{j\neq m} \braket{\sigma_j^+} \right).
\end{split}
\label{eq:squeezing_sigmaz}
\end{equation}

Next, we compute the contribution from the collective two-body dissipator
\begin{equation}
\begin{split}
    \frac{N^2}{\Gamma_2} \frac{d}{dt} \braket{\sigma_m^+} &= \sum_{i,j,k,l} \braket{ \sigma_i^+ \sigma_j^+ \sigma_m^+ \sigma_k^- \sigma_l^- - \frac{1}{2} \sigma_m^+ \sigma_i^+ \sigma_j^+ \sigma_k^- \sigma_l^- - \frac{1}{2} \sigma_i^+ \sigma_j^+ \sigma_k^- \sigma_l^- \sigma_m^+ } \\ &\approx \braket{\sigma_m^+ \sigma_m^-} \sum_{\substack{i,j,k \\ ijk \neq m}} \braket{\sigma_i^+ \sigma_j^+ \sigma_k^-} - \braket{\sigma_m^- \sigma_m^+} \sum_{\substack{i,j,k \\ ijk \neq m}} \braket{\sigma_i^+ \sigma_j^+ \sigma_k^-} - 2 \braket{\sigma_m^+ \sigma_m^- \sigma_m^+} \sum_{\substack{i,j \\ ij \neq m} } \braket{\sigma_i^+ \sigma_j^-} \\
    &= \braket{\sigma_m^z} \sum_{\substack{i,j,k \\ ijk \neq m}} \braket{\sigma_i^+ \sigma_j^+ \sigma_k^-} - 2 \braket{\sigma_m^+} \sum_{\substack{i,j \\ ij \neq m} } \braket{\sigma_i^+ \sigma_j^-}
\end{split}
\label{eq:two_photon_loss_sigma+}
\end{equation}
Eq. \eqref{eq:two_photon_loss_sigma+} contains terms such as $\braket{\sigma_i^+ \sigma_j^+ \sigma_k^-}$ and $\braket{\sigma_i^+ \sigma_j^-}$. To obtain a closed set of mean-field equations, we have to reduce these in terms of $\braket{\sigma_i^+}$ and $\braket{\sigma_i^z}$. Defining $c_{1,m} \equiv \frac{1}{N} \sum_{j \neq m} \braket{\sigma_j^+}$ and $c_{2,m} \equiv \frac{1}{2N} \sum_{j \neq m} (1 + \braket{\sigma_j^z})$, we then obtain
\begin{equation}
\begin{split}
    \frac{N^2}{\Gamma_2} \frac{d}{dt} \braket{\sigma_m^+} &\approx \braket{\sigma_m^z} \left( 2 \sum_{\substack{i,j \\ ij \neq m \\ i \neq j}} \braket{\sigma_i^+ \sigma_i^-} \braket{\sigma_j^+} + \sum_{\substack{i,j,k \\ ijk \neq m \\ i \neq j \neq m}} \braket{\sigma_i^+} \braket{\sigma_j^+} \braket{\sigma_k^-} \right) - 2\braket{\sigma_m^+} \left( \sum_{i\neq m} \braket{\sigma_i^+ \sigma_i^-} + \sum_{\substack{i,j \\ ij \neq m \\ i \neq j}} \braket{\sigma_i^+}\braket{\sigma_j^-} \right) \\
    &= -2 \braket{\sigma_m^+} (N c_{2,m} + N^2 |c_{1,m}|^2) + \braket{\sigma_m^z} (2N^2 c_{1,m} c_{2,m} + N^3 |c_{1,m}|^2 c_{1,m} )
\end{split}
\label{eq:two_photon_loss_sigma+2}
\end{equation}
Similarly, we can obtain the equation of motion for $\braket{\sigma_m^z}$:
\begin{equation}
\begin{split}
   \frac{1}{\Gamma_2} \frac{d}{dt} \braket{\sigma_m^z} &= \sum_{i,j,k,l} \braket{ \sigma_i^+ \sigma_j^+ \sigma_m^z \sigma_k^- \sigma_l^- - \frac{1}{2} \sigma_m^z \sigma_i^+ \sigma_j^+ \sigma_k^- \sigma_l^- - \frac{1}{2} \sigma_i^+ \sigma_j^+ \sigma_k^- \sigma_l^- \sigma_m^z } \\
   &\approx ( \braket{\sigma_m^+ \sigma_m^z} - \braket{\sigma_m^z \sigma_m^+} ) \sum_{\substack{i,j,k \\ ijk \neq m}} \braket{\sigma_i^+ \sigma_j^- \sigma_k^-} + (\braket{\sigma_m^z \sigma_m^-} - \braket{\sigma_m^- \sigma_m^z} \sum_{\substack{i,j,k \\ ijk \neq m}} \braket{\sigma_i^+ \sigma_j^+ \sigma_k^-} \\
   &+ (4 \braket{\sigma_m^+ \sigma_m^z \sigma_m^-} - 2 \braket{\sigma_m^z \sigma_m^+ \sigma_m^-} - 2 \braket{\sigma_m^+ \sigma_m^- \sigma_m^z} ) \sum_{\substack{i,j \\ ij \neq m}} \braket{\sigma_i^+ \sigma_j^-} \\
   &= -2 \braket{\sigma_m^+} \sum_{\substack{i,j,k \\ ijk \neq m}} \braket{\sigma_i^+ \sigma_j^- \sigma_k^-} - 2 \braket{\sigma_m^-} \sum_{\substack{i,j,k \\ ijk \neq m}} \braket{\sigma_i^+ \sigma_j^+ \sigma_k^-} - 4 (1 + \braket{\sigma_m^z}) \sum_{\substack{i,j \\ ij \neq m}} \braket{\sigma_i^+ \sigma_j^-} \\
   &= -4 \text{Re} \left( \braket{\sigma_m^+} \sum_{\substack{i,j,k \\ ijk \neq m}} \braket{\sigma_i^+ \sigma_j^- \sigma_k^-} \right) - 4 (1 + \braket{\sigma_m^z}) \sum_{\substack{i,j \\ ij \neq m}} \braket{\sigma_i^+ \sigma_j^-} \\
   &\approx -4 \text{Re} \left[ \braket{\sigma_m^+} \left( 2\sum_{\substack{i,j \\ ij \neq m \\ i \neq j}} \braket{\sigma_i^+ \sigma_i^-} \braket{\sigma_j^-} + \sum_{\substack{i,j,k \\ ijk \neq m \\ i \neq j \neq k}} \braket{\sigma_i^+}\braket{\sigma_j^-}\braket{\sigma_k^-} \right) \right] - 4 (1 + \braket{\sigma_m^z}) \left( \sum_{i\neq m} \braket{\sigma_i^+ \sigma_i^-} + \sum_{\substack{i,j \\ ij \neq m}} \braket{\sigma_i^+} \braket{\sigma_j^-} \right) \\
   &= -4(1+\braket{\sigma_m^z}) (N c_{2,m} + N^2 |c_{1,m}|^2 ) + \text{Re} [\braket{\sigma_m^z} (2 N^2 c_{1,m}^* c_{2,m} + N^3 |c_{1,m}|^2 c_{1,m}^*]
\end{split}
\label{eq:two_photon_loss_sigmaz}
\end{equation}

Combining Eqs. \eqref{eq:squeezing_sigma+}, \eqref{eq:squeezing_sigmaz}, \eqref{eq:two_photon_loss_sigma+2} and \eqref{eq:two_photon_loss_sigmaz}, we have the mean-field equations:
\begin{equation}
\begin{split}
    \frac{d}{dt}\braket{\sigma_m^+} &= i \omega_m \braket{\sigma_m^+} - 2i\eta e^{i\theta} \braket{\sigma_m^z} c_{1,m}^* - 2 \Gamma_2 \braket{\sigma_m^+} \left(\frac{c_{2,m}}{N} + |c_{1,m}|^2 \right) + \Gamma_2 \braket{\sigma_m^z} c_{1,m} (2c_{2,m} + N |c_{1,m}|^2 )
\end{split}
\end{equation}

\begin{equation}
\begin{split}
    \frac{d}{dt}\braket{\sigma_m^z} &= 8\eta \text{Im} \left( e^{-i\theta} \braket{\sigma_m^+}c_{1,m} \right) - 4\Gamma_2 (1+\braket{\sigma_m^z}) \left(\frac{c_{2,m}}{N} + |c_{1,m}|^2 \right) - 4\Gamma_2 \text{Re} [\braket{\sigma_m^+} c_{1,m}^* (2c_2 + N|c_{1,m}|^2 ) ]
\end{split}
\end{equation}
as described in Eqs.~\eqref{eq:mf_sigma+} and~\eqref{eq:mf_sigmaz} in the main text.

\section{Two detuned ensembles, small $\delta$ limit}
\label{app:twoensembles}
When the detuning $\delta$ is small, we expect the steady state value for the phase $\phi$ to be close to $\pi/4$, as was shown in Eq.~\eqref{eq:oneensemble_ss}. Writing $\phi = \pi/4 \pm \zeta$, we obtain the mean-field equations~(\ref{eq:twoensemble_MF1}),\,(\ref{eq:twoensemble_MF2}), and (\ref{eq:twoensemble_MF3}). Expanding the equations to linear order in $\zeta$, and then to order $1/N$, we obtain the simplifed equations (setting $\Gamma_2 = 1$ for notational simplicity)
\begin{equation}
\begin{split}
    \frac{\dot{A}}{A} &\approx - 2A^2 + \frac{N-1}{N} z (1-2 \eta + (N-2) A^2) + \frac{N-1}{N} z^2 \\
    \dot{\zeta} &\approx \delta - \frac{z}{N} (N + 4\eta - 2N \eta + (N-2) A^2 + N z) \zeta \\
    \frac{\dot{z}}{A^2} &\approx \frac{4A^2}{N}(1+2N(\eta-1) - 2\eta - (N-2)(N-1) A^2 + z - 2Nz).
\end{split}
\end{equation}
Since $A^2$ is of order $1/N$, one has to be careful here when making the large $N$ approximation. We verify the accuracy of the above equation by comparing numerical simulations against those from the full mean-field equations~\eqref{eq:mf_sigma+} and~\eqref{eq:mf_sigmaz}. The steady state solution is complicated, but we can expand in powers of $1/N$ to get
\begin{equation}
    A^2 \approx \frac{2\eta}{N} - \frac{4\eta}{N^2} + \frac{4\eta - 32\eta^2}{N^2} + \ldots \approx \frac{2\eta}{N},
\end{equation}
\begin{equation}
    \zeta \approx \delta \left( \frac{1 + \sqrt{1-16(\eta/N)}}{32 (\eta/N)^2} + \frac{1+32(\eta/N)}{128N(\eta/N)^3} + \ldots \right) \approx \frac{1 + \sqrt{1 - 16 \eta/N}}{32 (\eta/N)^2} \delta
\end{equation}
and 
\begin{equation}
    z \approx - \frac{1 + \sqrt{1-16\eta/N}}{2}
\end{equation}

\end{document}